\documentclass{pasa}%

\usepackage{aas_macros}
\usepackage[usenames,dvipsnames,table]{xcolor}
\usepackage{hyperref}
	\hypersetup{	linktoc=page,
				colorlinks=true,
				linkcolor=Maroon,
				citecolor=PineGreen,
				urlcolor=RoyalBlue}
\usepackage{amssymb}
\usepackage{wasysym} 
\usepackage{cosmotix,minitix} %KM Cosmology and formatting tools
\usepackage{ulem} %For editing for resubmission

\title[Lyman-$\alpha$ Forest and Cosmic Weak Lensing in a Warm Dark Matter Universe]{Lyman-$\alpha$ Forest and Cosmic Weak Lensing in a Warm Dark Matter Universe}
\author[Markovi\v{c}, K. and Viel, M.]{Katarina Markovi\v{c}$^1$\thanks{Email: 
	\href{mailto:markovic@usm.lmu.de}{markovic@usm.lmu.de}}
	\and Matteo Viel$^{2,3}$ \thanks{Email: 
	\href{mailto:viel@oats.inaf.it}{viel@oats.inaf.it}} \\
\affil{$^1$University Observatory Munich, Ludwig-Maximilian University, Scheinerstr. 1, 81679 Munich, DE}%
\affil{$^2$INAF - Osservatorio Astronomico di Trieste, Via G. B. Tiepolo 11, I-34143 Trieste, IT}\affil{$^3$INFN/National Institute for Nuclear Physics, Via Valerio 2, I-34127 Trieste, IT}}%
\jid{PASA}
\doi{10.1017/pas.\the\year.xxx}
\jyear{\the\year}

\usepackage[authoryear]{natbib}
	\bibpunct{(}{)}{;}{a}{}{,}
\renewcommand{\redtxt}[1]{#1}

\begin{document}%
\begin{abstract}
We review the current state of the theory of large scale structure in a warm dark matter (WDM) cosmological model. In particular, we focus on the non-linear modelling of the matter power spectrum and on the mass function of dark matter haloes.  We describe the results of N-body simulations with WDM and mention the effects that could be induced by baryonic physics. We also examine the halo model of large scale structure and its recently suggested modifications for a WDM cosmology, which account for the small scale smoothness of the initial matter density field and better fit the results of N-body simulations. Having described the theoretical models, we discuss the current lower limits on the WDM particle mass, $m_\WDM$, which correspond to upper limits on the WDM temperature under the assumption that the particles are thermal relics. The best such constraints come from the \Lya forest and exclude all masses below $3.3\keV$ at the $2\sigma$ confidence level. We finally review the forecasts for future lensing surveys, which will be of the same order of magnitude as the already existing constraints from the \Lya forest data but explore a different redshift regime.\end{abstract}
\maketitle%
%

%%%%%%%%%%%%%%%%%%%%%%%%%%%%%%%%%%%%%%%
%%%%%%%%%%%%%%%%%%%%%%%%%%%%%%%%%%%%%%%
\section{Introduction}

Here, we briefly outline how the idea of Dark Matter (DM) emerged and
when Warm Dark Matter (WDM) branched off the mainstream theory.  The
idea that the temperature of DM impacts the distribution of structure
in the universe is as old as the notion that galaxies cluster on large scales.

An important step in dark matter history, which started with the
measurements of displacements within spiral galaxies to measure their
rotation, was to realise that the dynamical properties of galaxies and
galaxy clusters did not seem to match their observed luminous mass
\citep[see e.g.][and references
  therein]{Zwicky:1937,Schwarzschild:1954,Janak:1958,Abell:1962,Burbidge:1969,Paal:1976}. Moreover,
thanks to the first Palomar Observatory Sky Survey
\citep[1949-1958,][]{Reid:1991}, the distribution of galaxies in the
sky was mapped for the first time in the mid 20\th century and showed
that galaxies conglomerated not only into clusters, but also gave rise
to the cosmological Large Scale Structure
\citep{Press:1974,Rudnicki:1976,White:1978,Jones:1978,Wesson:1978,Aarseth:1980},
whose properties depended on cosmological parameters.

It thus became clear early on that the so-called ``missing mass
problem'' \citep[e.g.][]{Faber:1979} was unlikely to be solved by dark
gaseous matter, that it had to be dark stars, black holes, comets or
something else, like the conveniently weakly interacting massive
particle - the neutrino. Assuming the missing mass was made up of
massive neutrinos and other weakly interacting particles, an upper
limit of 
\redtxt{$8\eV/c^2$}
could be placed on their masses, assuming
the measured 
\redtxt{expansion of the universe}\footnote{\redtxt{This was consistent with the measurement of neutrino mass from the Coma cluster density profile of \citet{Cowsik:1973}.}}
\citep{Cowsik:1972}.

The discovery of the Cosmic Microwave Background (CMB) by
\citet{Penzias:1965,Dicke:1965} resulted in the acceptance of the Hot
Big Bang origin of the universe. Big Bang Nucleosynthesis (BBN) put
severe constraints on the amount of baryonic matter in the universe
\citep[$\Omega_\b \ll 1$,][and references
  therein]{Schramm:1977,Olive:2000}, and, combined with the
requirement that the total density in the universe was close to the
critical density for flatness\footnote{Flatness implies that the
  energy density in the universe is equal to the ``critical density'' for
  flatness, i.e. 
  \redtxt{the total density parameter,}
  $\Omega\sim1$, measured from the Hubble parameter and
  assumed from ``naturalness'' of 
  \redtxt{zero spatial curvature, $\kappa=0$} 
  of the Einstein-de Sitter
  model.}, gradually led to the belief that DM is made of a new
elementary particles.

One of the first candidates was a massive neutrino, but more general
new particles were considered quickly, like other stable neutral
leptons \citep{Gunn:1978,Tremaine:1979}. The theory of Supersymmety
(SUSY) offered new candidate particles, like the \textit{gravitino}
with a $\keV$ mass \citep{Pagels:1982}, or the \textit{photino}
\citep{Sciama:1983}. In addition the \textit{axion}, whose Jeans Mass
would be smaller than galactic scales, was considered as a DM
candidate \citep{Stecker:1983,Shafi:1984}.  These particles were
distributed in the universe as a perturbed density field, which became
gravitationally coupled to the density field of baryons in the matter
dominated era. However, because there was no energy dissipation 
\redtxt{for the particles of DM, they}
could not condense into bound objects easily. This
was especially true for massive neutrinos with masses $\sim\eV$, whose
damping scales reached the sizes of galaxy clusters and even
superclusters \citep{Bond:1980,Schramm:1981}. This meant that if the
DM that was closing the universe was made up of massive neutrinos, the
distributions of dark and luminous matter would have to be very
different \citep{Bruns:1983}. In fact, in such a neutrino-dominated
Hot Dark Matter (HDM) model, the mechanism for galaxy formation was
considered to be a fragmentation of large objects, which collapse
first, as proposed by \citet{Zeldovich:1970}. These objects, which
collapsed along one dimension first, were known as ``pancakes''. After
collapsing along two dimensions, they became filaments and finally
\redtxt{spherically symmetric}
DM \textit{haloes}.

The observations of the dynamics of galaxies, galaxy clusters and
superclusters were compared to the amount of visible mass by many
authors \citep[e.g.][and references therein]{Bruns:1983}. A trend
seemed to emerge, where the ``missing mass'' fraction increased with
object size, implying that the relation between the distributions of
dark and luminous matter densities must be more biased on large
scales.

Simultaneously, hierarchical merging of structure was considered via
N-body simulations of the formation of the large scale structure in
the universe \citep{Aarseth:1980,White:1982}. In this picture,
structure formed as a consequence of pure gravitational collapse of
the initial linear density perturbations. Such ``bottom-up''
hierarchical structure formation occurred if the matter density in the
universe was dominated by particles more massive than at least several
tens of $\eV$ \citep{Bond:1982}, but it was not possible for the
$\m_\nu\sim 30\eV$ neutrinos \citep{Peebles:1982}. The bottom-up
scenario became strongly favoured in 1984, since observations of dwarf
galaxies as well as
\redtxt{those of large scale structure overall}
put strong lower limits on the mass
of the DM particle \citep{Lin:1983,Kaiser:1983,Madsen:1984}. In other
words, the standard model of DM became the Cold Dark Matter (CDM)
model.\\

\redtxt{However,} 
\citet{Klypin:1999} noticed a discrepancy in the observed numbers of
the smallest galaxies, assumed to reside within DM haloes with masses
$M_{\rm dwarf} \leq 10^{9} M_\odot$ and what they expected these
numbers to be from running their numerical simulations of structure
formation
\redtxt{with CDM}.
They proposed that the numerical simulations might be
modified to account for this discrepancy, which has become known as
the \textit{dwarf galaxy problem} or \textit{missing satellite
  problem} \citep{Bode:2000}, \redtxt{since the number of small
  objects observed fell significantly short of the expectation.}

In recent years, N-body and hydro-dynamical codes have improved
significantly in matching the small scales of \LCDM to
  observations, for example via the suppression of the formation of
  baryonic objects within small DM haloes \citep[e.g. ``cosmic web
    stripping'',][]{Benitez:2012}.  In addition, surveys like the
Sloan Digital Sky Survey have found new dwarf galaxies around the
Milky Way \citep[see][and references within for a review of the
  problem]{Bullock:2010}.

It is possible that 
\redtxt{this}
small-scale crisis of \LCDM could be solved or
alleviated with better numerical prescriptions for the complex
baryonic processes \citep[see e.g.][]{Brooks:2012} or it could be due
to observational biases. However, the density profiles and
concentrations of individual haloes \citep{Donato:2009} as well as the
properties of voids \citep{Tikhonov:2009} do not seem to match what
one would expect from pure \LCDM model. Baryonic processes are
difficult to invoke for explaining the properties of dwarf galaxies
that are dark matter dominated, making it hard to fit their
observational properties.

These long known ``missing satellite'' and ``core-cusp problems'' are
related to more recently defined ``too-big-to fail'' problem, being
that the most massive Milky Way subhaloes from local \LCDM simulations
do not have dynamical properties similar to the observed Milky Way
dwarf galaxies. For a recent review of the small scale issues of
\LCDM see \citet{Weinberg:2013}.

An elegant solution \redtxt{has been considered in the past, which has
  introduced} WDM in a simple \LWDM model with one additional
parameter, which could explain all or some of these
discrepancies. Because of its free-streaming, WDM is capable of
damping the density field on small scales without any change to the
large scale behaviour of structure or to the dynamical evolution of
space-time. For this reason we now \redtxt{discuss and summarise} how
to calculate non-linear corrections to predict the statistical
properties of cosmological structure. This is not a straightforward
task, but nonetheless, we describe attempts to develop a prescription
valid also in \LWDM models that may be used one day to account for
\redtxt{some of} the discrepancies at small scales of \LCDM.

\redtxt{However, it should be noted that there exists a phase-space density imposed
  lower bound on the fermionic DM particle mass, called the
  Tremaine-Gunn bound \citep{Tremaine:1979,Hogan:1999}, due to which
  it may not be possible for a WDM model to produce the relatively
  large cores that we seem to observe. Large halo cores can namely
  only be produced by very small particle mass, $m_\WDM$
  \citep{Shao:2013}.}  This has become known as the
``too-small-to-succeed'' problem and put significant pressure on the
WDM scenario.\\

In this review, we discuss \redtxt{the existing literature on}
constraining the WDM particle mass, $m_\WDM$ using the statistical
properties of the large scale structure. We choose this approach in
hope that it may contain some information not contaminated by the
uncertainties arising out of a lack of a rigorous model of baryonic
feedback and cooling processes. There are many other works that look
at individual objects of the large scale structure and hope to
constrain WDM from their properties
\citep[e.g.][]{Pacucci:2013,Lovell:2011,Maccio:2012,Vinas:2012,SommerLarsen:2001}. 

In particular, we discuss in some detail the modelling of non-linear
large scale structure needed for comparison with data. We choose two
observables to constrain our models: the \Lya forest and the cosmic
shear (weak lensing) power spectrum, both of which require an accurate
modelling of the non-linear matter power as a first step in the
modelling of their basic properties.  In \autoref{sec:frame} we
summarise the general background physics of the smoothing of the
linear matter density field by the free-streaming dark matter and the
calculation of the linear matter power spectrum. We also briefly
discuss the particle candidates for WDM. We then describe
prescriptions for calculating the non-linear matter power spectra in
the WDM scenario. In particular, we discuss N-body simulations, the
halo model and the current status of the two approaches for
calculating the statistics of the large scales in the universe. We
summarise the current limits coming from the \Lya forest data in
\autoref{sec:meas}, which present the strongest constraints on the
temperature of dark matter to date. Finally, we report on forecasts
that have been made on the WDM temperature obtained from future weak
lensing surveys like Euclid\footnote{\cite{Amendola:2012,Refregier:2010}} in \autoref{sec:lens}.

%%%%%%%%%%%%%%%%%%%%%%%%%%%%%%%%%%%%%%%
%%%%%%%%%%%%%%%%%%%%%%%%%%%%%%%%%%%%%%%
\section{General Framework}\label{sec:frame}

Neutrinos decouple when the temperature of the primordial plasma is
$T\sim1\MeV$ and $a\sim10^{-10}$ and become non-relativistic when
$T_\h \sim m_\h/3k_{\rm B}$\footnote{The Boltzmann constant, $k_{\rm
    B}= 8.617\times10^{-5}\eV\ {\rm K}^{-1}$.}. DM decouples and
becomes non-relativistic much earlier in both the CDM and WDM
cases. If WDM has a simple thermal history, analogous to neutrinos,
but with a larger particle mass, we can calculate its
\textit{free-streaming}. Such a DM particle is called a
\textit{thermal relic}, \redtxt{because it was once in equilibrium
  with itself.}

The Jeans length can be calculated for a perfect fluid and denotes the
limit on which the gravitational effect balance out the thermal
effects \citep{Bond:1983}. For collisionless fluids like the DM and
neutrino fluids, we can define the analogous comoving free-streaming 
\redtxt{wavenumber},
which tells how far the fast-moving particles can travel within the
gravitational time-scale i.e. in the time of free-fall \citep{Boyarsky:2008}:
\ba \label{eq:kfsv}
k_\fs(a) &=& \sqrt{\frac{3}{2}}\frac{aH(a)}{v_{\x,{\rm median}}} \comma 
\ea 
where $v_\x=1$, when the particles are relativistic. When they go
non-relativistic (i.e. when $3k_{\rm B}T_{0,\x}\lesssim m_\x$), 
\be
v_\x = \frac{3k_{\rm B}T_{0,\x}}{a m_\x} 
\ee 
and then $k_\fs
\rightarrow \infty$ as $a \rightarrow 1$ and $T \rightarrow 0$, which
is the case for CDM very early on, and therefore, the effects of
free-streaming are pushed to very very large $k$, i.e. extremely small
scales. This means that the damping of the overdensity field becomes
insignificant at cosmological scales.

In fact, even in mixed DM models \redtxt{(C+HDM)}, where the HDM
component makes up a small fraction of the total energy density as in
the \LCDM + neutrinos, $\nu$\LCDM, the perturbations in the cold
component are modified by the free-streaming of the HDM. In this
scenario the larger scales suffer more free-streaming damping and
therefore the perturbations in the HDM cannot grow until late times,
which gravitationally affects the perturbations in the cold component,
slowing down the growth of the perturbation amplitudes
\citep{Primack:1998,Ghigna:1997,Klypin:1995,Nolthenius:1994,Klypin:1993,Gawiser:1998,Primack:1997,Primack:1995,Ma:1995,Zentner:2003,Primack:2003}.

The most basic model of WDM particles is to assume that they are thermal
relics. This means that they were in
\redtxt{thermal}
equilibrium at some point. When
their temperature and density dropped, they went out of equilibrium
\citep[e.g.][]{Bond:1983}  and DM particles decoupled from
each other. Instead, the sterile neutrino particles, that will be
discussed later, were never in thermal equilibrium.

Theoretically there would have been another kind of decoupling. This
would have been when DM particles and baryons were in an extremely
dense environment and so there would have been a significant
interaction rate between them. We know very little about this regime,
because we would have to know the mass and interaction cross-section
of DM particles, but we do not even know the nature of the interaction
(if any) between DM particles and other types of matter.

However, it is most likely that these two decouplings happened at the
same time, because any self-interaction of DM is likely to involve the
weak, strong or electromagnetic force, which means this
self-interaction would necessarily involve baryons. Were this not the
case, it may be that the interaction between baryons and DM particles
is weaker than the interaction among DM particles. In this case the
decoupling from baryons would happen at an earlier time than
decoupling of DM out of equilibrium.

The smallest important scale feature in the linear matter power
spectrum is the suppression induced by DM free-streaming. In the WDM
model the scale of suppression is called the free-streaming scale,
$k_{\rm fs}$ and corresponds to the mode that enters the horizon at
the time when WDM particles become non-relativistic, $t_{\rm rel}$. A
species can become non-relativistic while still in thermal equilibrium
or after it decouples \citep{Bond:1983,Bode:2000,White:1987}. If it is
after, we say that DM particles decouple while non-relativistic. This
is what we often assume in modelling the large scale structure in
these models, for the sake of simplicity.

From \citet{Bond:1983}, we get the temperature of WDM relative to that of the photons,
from which one can calculate the total WDM density for a particular particle model (giving $g_\WDM$ and $m_\WDM$):
\be \label{eq:wdmtemp}
\Omega_\WDM = \frac{1.1}{h^{2}}\left(\frac{100}{3.9}\right)\left(\frac{g_\WDM}{1.5}\right)
	\left(\frac{T_\WDM}{T_\gamma}\right)^3\left(\frac{m_\WDM}{1 \keV}\right) 
\tab ,
\ee
where $g_* = 3.9 \left(T_\gamma/T_\WDM\right)^3$ is the number of all relativistic {degrees of freedom} at WDM decoupling, $g_\WDM$ are the degrees of freedom for the WDM, $T_\gamma$ is the present day photon temperature and $T_\WDM$ is the temperature of WDM. We can calculate the degrees of freedom:
\be
g_\WDM  = \left\{
	\begin{array}{l l}
		N_\WDM & {\rm bosons} \\
		\frac{3}{4}N_\WDM & {\rm fermions} \tab,
	\end{array} \right.
\ee
where $N_\WDM$ are the number of spin degrees of freedom. Then assuming $\Omega_\DM=\Omega_\WDM$ gives a direct relationship between $T_\WDM$ and $m_\WDM$. Otherwise we must introduce a new parameter $f_\WDM = \Omega_\WDM/\Omega_\DM$, the fraction of WDM. This parameter becomes relevant when we start to consider C+WDM models.
% MDM is mixed dark matter as above, changing this to C+WDM everywhere.

In addition we can calculate the \textit{velocity dispersion} of WDM particles relative to that of the neutrinos \citep{Bond:1980}:
\be
\sqrt{\left< v^2\right>_\nu} 
	= 6 \kms \left(\frac{30\eV}{m_\nu}\right)(1+z) \tab,
\ee
Rescaling for WDM, if it has decoupled while relativistic:
\be
\sqrt{\left< v^2\right>_\WDM} = 
	\sqrt{\left< v^2\right>_\nu}\left(\frac{T_\WDM}{m_\WDM}\right)
	\left(\frac{m_\nu}{T_\nu}\right)  \tab .
\ee
If particles decouple while non-relativistic, $\sqrt{\left< v^2\right>_\WDM} \lesssim \cms$, so 
\redtxt{any further damping is insignificant and the species becomes effectively ``cold''.}

%%%%%%%%%%%%%%%%%%%%%%%%%%%%%%%%%%%%%%%
\subsection{The Linear Power Spectrum}\label{sec:lin}

\begin{figure*}[!ht]
\centering
\includegraphics[width=0.502\textwidth,clip=]{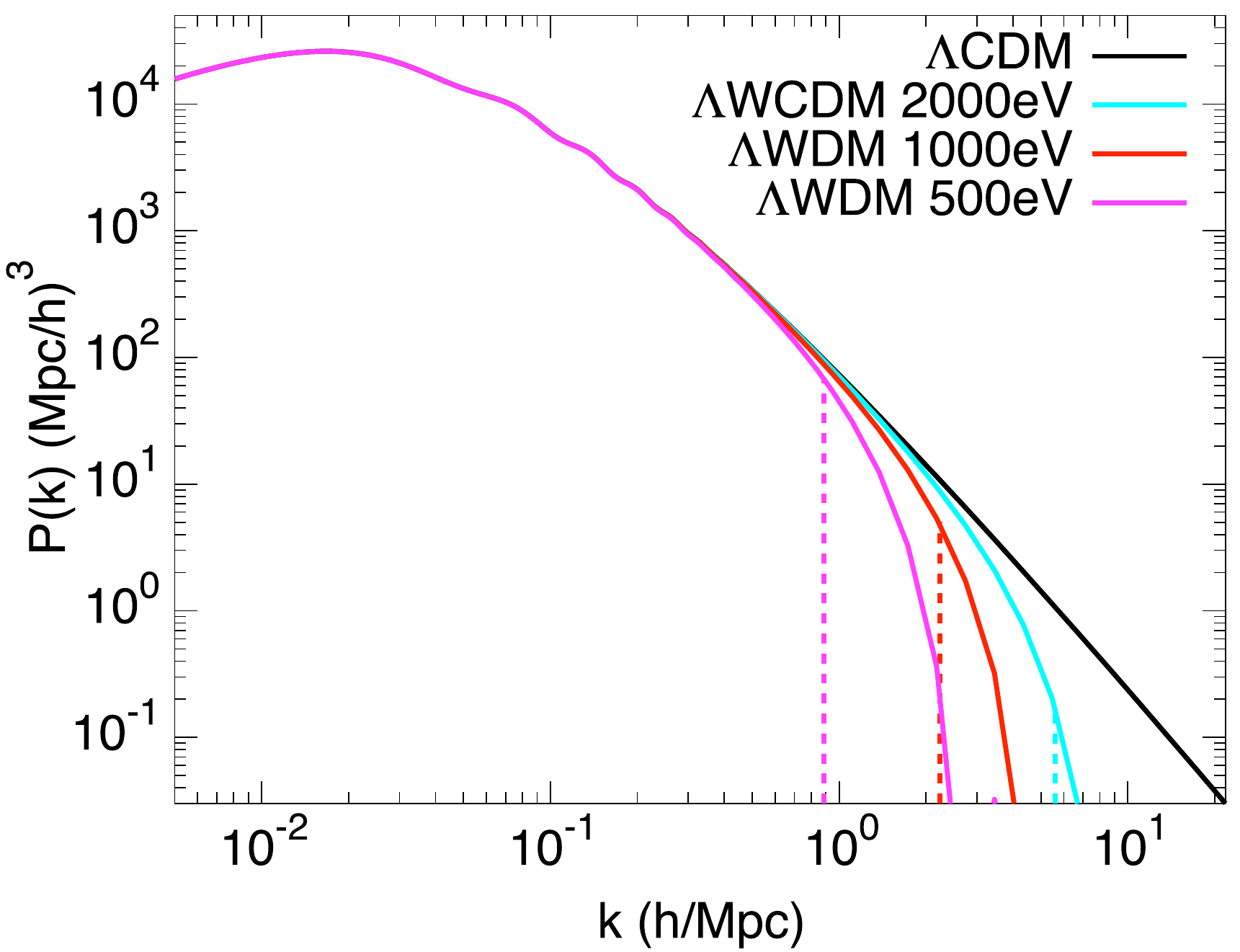}
\includegraphics[width=0.488\textwidth,clip=]{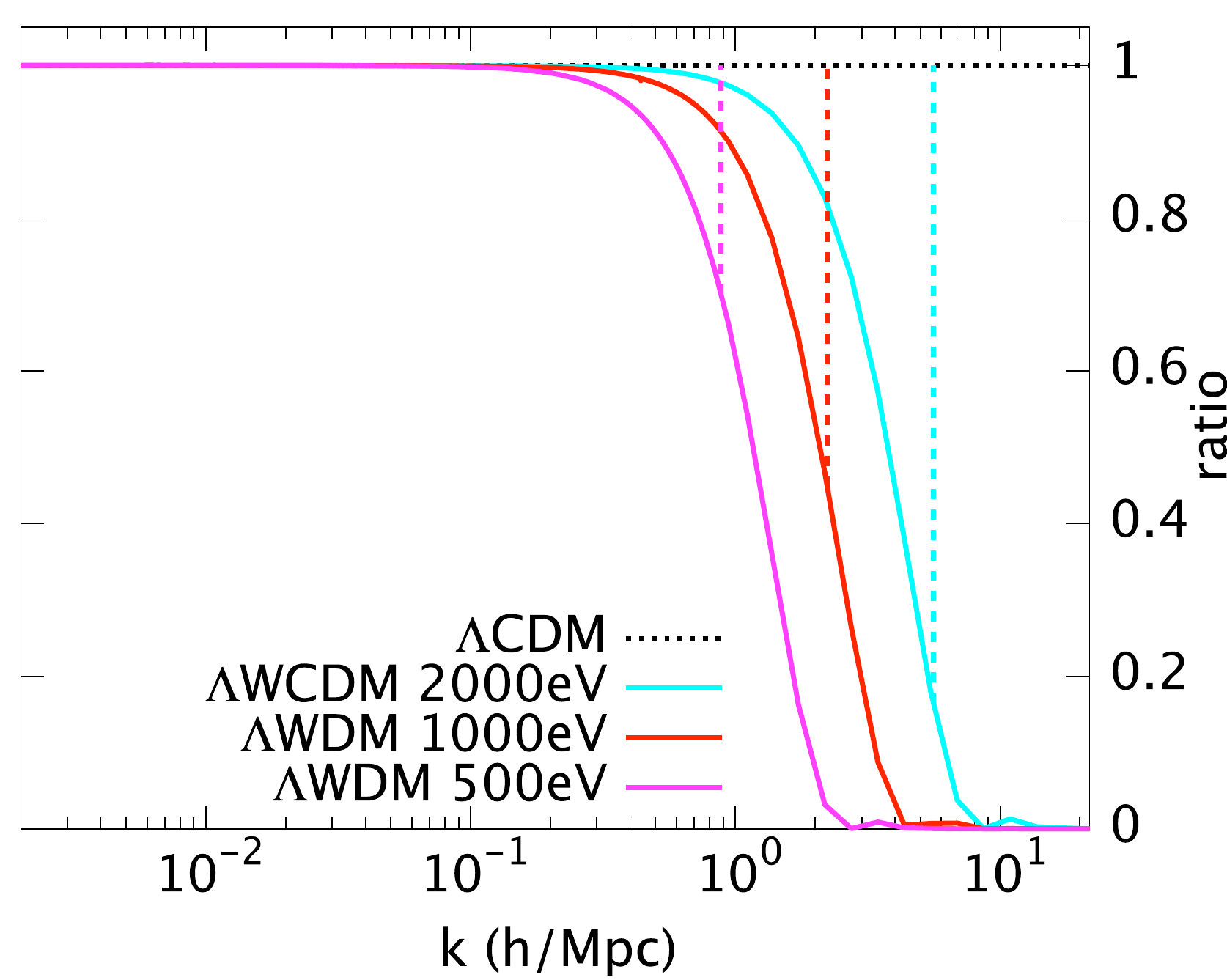}
\caption{\textit{Left}: The linear matter power spectra for three different WDM models and standard CDM. The particle masses, $m_\WDM \in \{0.5,1.0,2.0\} \keV$ are colour coded with magenta, red and cyan, respectively. The vertical lines correspond to a tenth of the free-streaming wavenumber, $k_\fs/10$, for each model of WDM. These power spectra were produced using the Boltzmann solver \code{CLASS} \citep{Lesgourgues:2011}.\newline
\textit{Right}: ratios between the WDM and CDM power spectra, $P_\WDM(k)/P_\CDM(k)$, to clearly see the suppression with respect to the $k_\fs/10$.} \label{fig:linear}
\end{figure*}
In the case of WDM, the initial matter power spectrum emerging from radiation domination is modified by an additional transfer function due to free streaming. \citet{Viel:2005} used a fitting function that can however be calculated very accurately with a numerical Boltzmann equation solver code, like for example \code{CMBFAST} \citep{Seljak:1996}, \code{CAMB} \citep{Lewis:2000} or 
\code{CLASS} \citep{Blas:2011,Lesgourgues:2011,Lesgourgues:2011b}. These codes solve the equations describing the growth of perturbations in the radiation dominated universe in a semi-analytic line-of-sight approach.

The fitting function of \cite{Viel:2005} with $\nu=1.12$ \citep[the alternative is $\nu=1.2$ like in ][]{Bode:2000} contains a scale-break parameter,
which is used in calculating the linear matter power spectrum by multiplying with the following WDM transfer function:
\ba \label{eq:linearpswdm}
\mathcal{T}_\WDM(k) &=& \left(1+(\alpha k)^{2\nu}\right)^{-5/\nu} 
	\tab \textrm{and so} \\
P^\lin_\WDM(k) &=& P^\lin_\CDM(k) \mathcal{T}^2_\WDM(k) \comma \nonumber
\ea
\redtxt{where the scale breaks at:}
\be \label{eq:scalebreak}
\alpha = 0.049\left(\frac{m_\WDM}{1 \keV}\right)^{-1.11}
	\left(\frac{\Omega_\WDM}{0.25}\right)^{0.11}\left(\frac{h}{0.7}\right)^{1.22} \fullstop
\ee
Alternatively, \cite{Boyanovsky:2008} found a transfer function for a general initial thermal distribution of DM particles - cold WIMP dark matter, thermal fermionic or bosonic dark matter.\\
The linear power spectrum, $P^\lin_\WDM(k)$, must then be normalised to ensure the value $\sigma_8^2$ at 
\redtxt{$R = 8 \hMpc$.}
Finally we now can plot the linear matter power spectra in \autoreft{fig:linear}. The lightest WDM particle mass shown (500 eV) causes the linear theory matter power spectrum to be suppressed dramatically at a wavenumber significantly above 1 \ihMpc. The matter power spectrum of WDM starts to turn off well above the free-streaming scale, which changes the slope of the power spectrum to fall much more steeply than $n_{\rm eff}=\log P(k)/\log k=-3$, which is the slope for standard, bottom-up structure formation \citep{White:1991,Knebe:2003}. 

In fact the above seemingly artificial scale break, $\alpha$ relates
to the free-streaming length of thermal relic WDM particles
\citep{Zentner:2003}: \be \label{eq:fslength} \lambda_{\rm fs} \simeq
0.11 \left[\frac{\Omega_\WDM h^2}{0.15} \right]^{1/3}
\left[\frac{m_\WDM}{\keV}\right]^{-4/3} {\rm Mpc} \tab , \ee which of
course is related to the \redtxt{Fourier space free-streaming scale,
  where free-streaming length effect contribute most to the power (see
  also \autoref{eq:kfsv})}: \be k_\fs \sim 5 \Mpc
\left(\frac{m_\WDM}{\keV}\right)\left(\frac{T_\nu}{T_\WDM}\right) \tab
.  \ee We plot $k_\fs/10$ in \autoreft{fig:linear}, because it is the
scale around which the significant suppression of the power in the
linear matter power spectrum begins. It is an interesting
\redtxt{open-ended} question why the free-streaming suppression
reaches scales so much larger than the free-streaming scale. This has
been explored, among others, by \citet{Smith:2011}.

In addition, we can define a corresponding mass found, on average, in
a volume with such a radius or \textit{free-streaming mass}:
\be \label{eq:fsmass} M_\fs = \frac{4\pi\rho_{\rm
    m}}{3}\left(\frac{\lambda_\fs}{2}\right)^3 \tab .  \ee where
$\rho_{\rm m}$ is the comoving background matter density and
$\lambda_{\rm fs}$ is the comoving free-streaming length defined in
\autoreft{eq:fslength} (note that different definitions for the free
streaming mass are used in the literature). We  will come back on this issue in \autoref{sec:halo}.

\begin{figure*}
	\begin{centering}
		\includegraphics[width=0.502\textwidth]{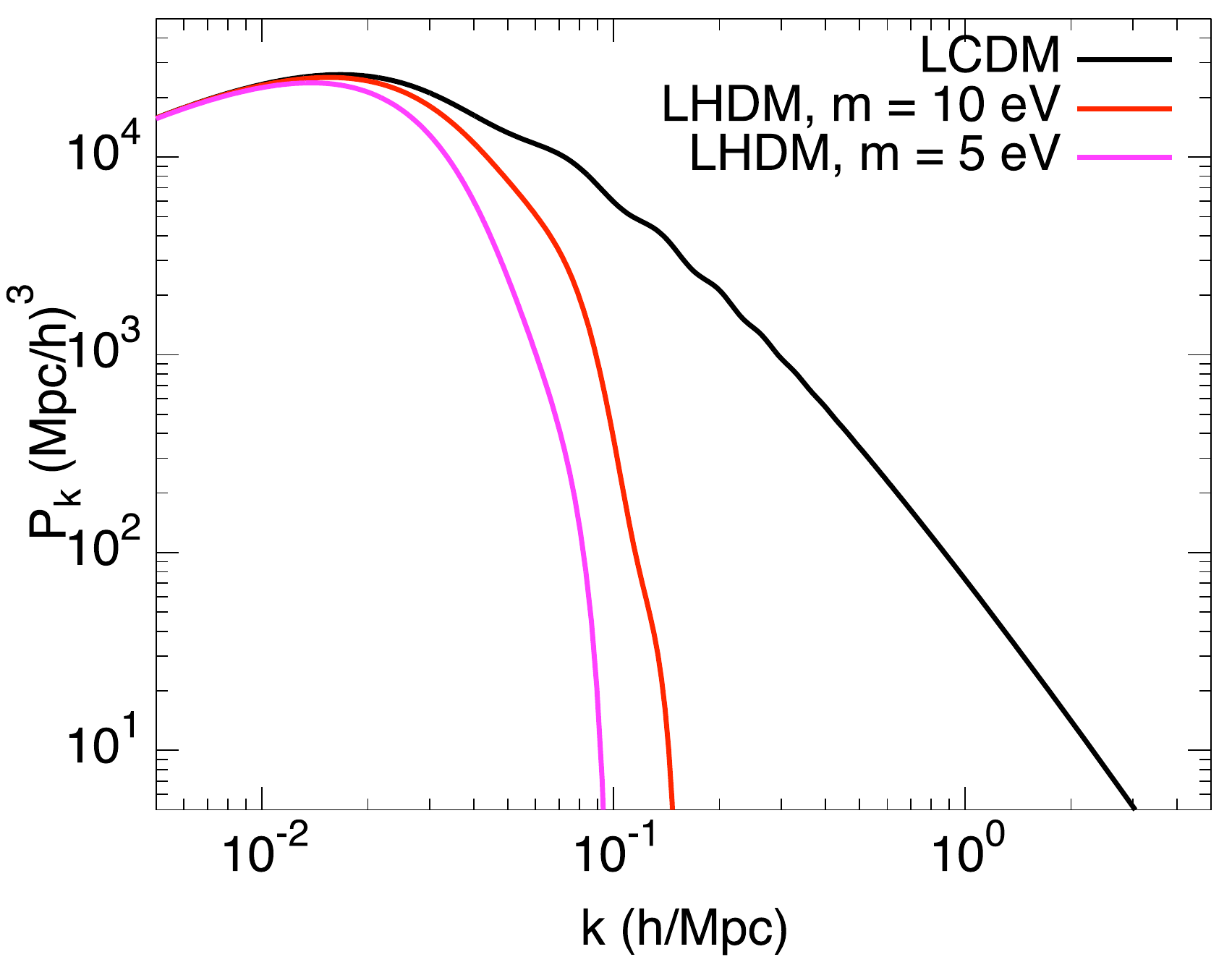}
		\includegraphics[width=0.49\textwidth]{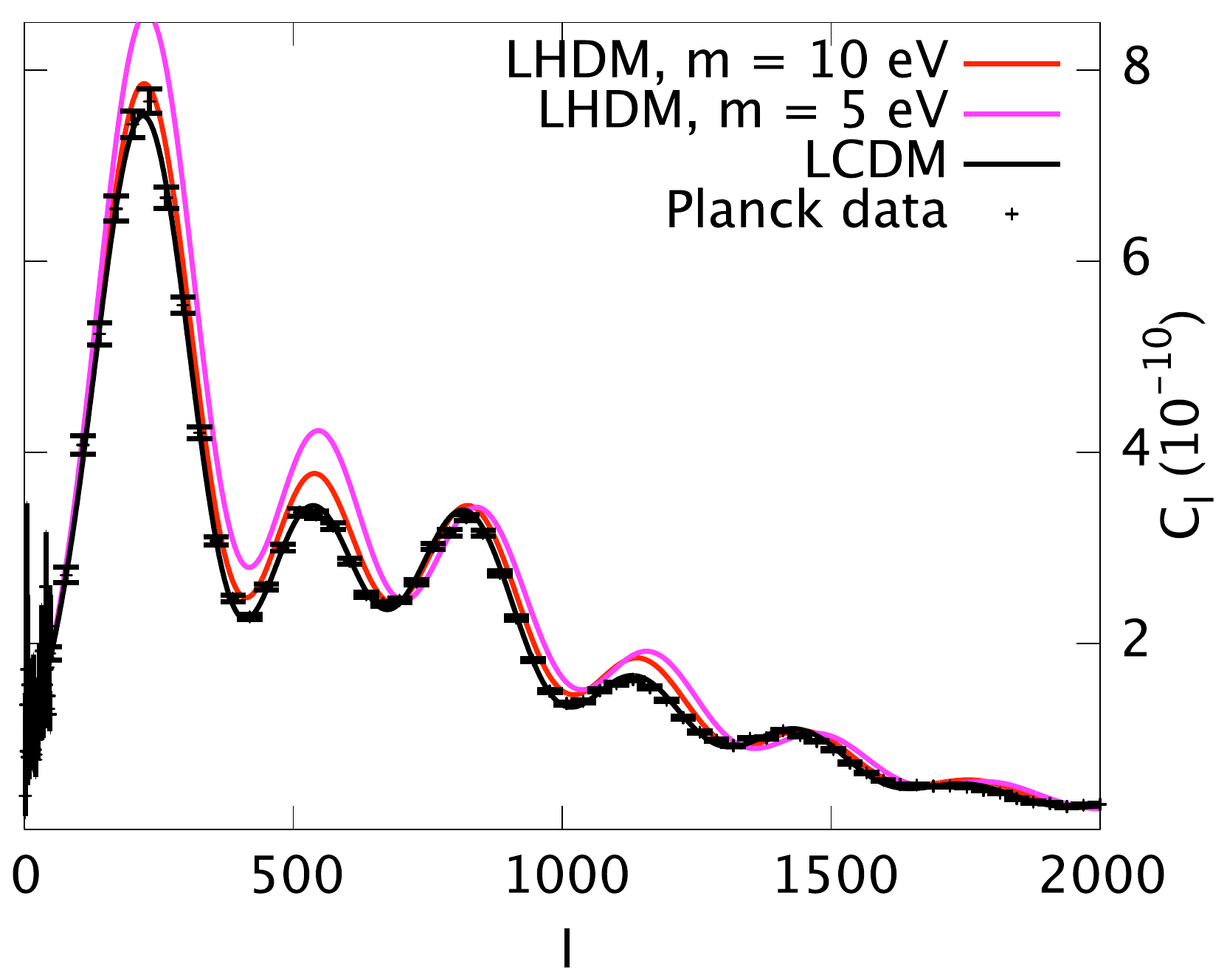}
		\caption{We plot the effect of a hot thermal relic
                  particle (hotter than a WDM candidate) on the Cosmic
                  Microwave Background $C(l)$'s: we show models that
                  have been long ruled out, where all the DM is made
                  up of very light, neutrino-like particles with
                  masses of $m_\WDM = 0.005$ and $0.01\keV$.\newline
                  \textit{Left:} The 3D linear matter power spectrum,
                  $P(k)$ for CDM (black) and the two ``HDM'' models
                  (magenta and red, respectively). Compare this extreme
                  case to the more plausible models in
                  \autoreft{fig:linear} that will impact at a much
                  smaller level. \newline \textit{Right:} The CMB
                  un-lensed $C(l)$'s in the corresponding HDM models.
                  The cosmological model is the Planck best fit as in
                  the rest of the paper, the error bars are those of
                  Planck survey \citep{Planck:2013}.  }\label{fig:hdm}
	\end{centering}
\end{figure*}
In \autoreft{fig:hdm} we plot the linear matter power spectrum
alongside the CMB power spectrum for the HDM scenario to illustrate
the impact on the CMB and matter power of such a small mass candidate, it is thus clear that heavier
masses will impact much less on these two observables at the largest scales. 
We plot the power spectra in the neutrino-like (but with $m\sim10\eV$) scenario being all of DM. 
\redtxt{We expect the power spectra to be suppressed at very large $l$ due to the free-streaming effect and to increase for small $l$, due to a mixture of two effects: changes in the matter-radiation equality and because the primordial power spectrum is normalised at $k = 0.05 \ihMpc$ causing the small-scale-suppressed power spectrum to be boosted on large scales.}

Since, $l=2000$ corresponds to about a $k\sim 0.2\, h/$Mpc at $z_{\rm CMB}=1000$, we expect the effect on the CMB from reasonable WDM scenarios to be completely negligible. In the right panel of \autoreft{fig:linear} it can be noted that at $k=0.2, h/$Mpc, the suppression is clearly less than 1\% for $m_\WDM = 1\keV$.

%%%%%%%%%%%%%%%%%%%%%%%%%%%%%%%
\subsection{Sterile Neutrinos}

We have discussed some of the particle candidates for DM in this
introductory section of this review. A further hypothetical particle
that has sparked interest is the \textit{sterile neutrino}, which does
not require an extension of the Standard Model with SUSY
\citep{Dodelson:1994,Fuller:2003,Asaka:2005,Abazajian:2005,Boyarsky:2006,Petraki:2008,Laine:2008,Kusenko:2009,Hamann:2010,Boyarsky:2013}. Sterile
neutrinos are assumed to be singlet right-handed particles that become
relatively heavy compared to standard, active neutrinos, which receive
small masses via a see-saw mechanism \citep{Dodelson:1994}. The
lightest of the additional neutrinos can then have a mass in the \keV
range, meaning that it resembles a WDM. However, the sterile neutrinos
are assumed to never have been in thermal equilibrium, therefore,
\redtxt{their} abundance was suppressed.

In the above calculation of the free-streaming scale we have used
three parameters that describe the thermal relic WDM model: the
particle mass, $m_\WDM$, the energy density, $\Omega_\WDM$ and the
temperature $T_\WDM$, where the degrees of freedom at WDM decoupling,
$g^*_\WDM$ are determined solely by the particle mass and its energy
density. We can conveniently re-parametrise the particle mass of the
never-thermalised sterile neutrino in terms of the thermal relic mass,
such that they are interchangeable in the calculation of the impact of
their free-streaming on the large scale structure:
\begin{equation} \label{eq:sterilemass}
  m_{\nu{\rm s}} = 4.43 \left(\frac{m_\WDM}{{\rm keV}}\right)^{4/3} \left(\frac{\omega_\WDM}{0.1225}\right)^{-1/3}{\rm keV} \comma
  \end{equation}
where $\omega_\WDM=\Omega_\WDM h^2$ \citep{Viel:2005}\footnote{with
  $h=H_0/100$ ${\rm km}$ ${\rm s}^{-1}{\rm Mpc}^{-1}$, the Hubble
  parameter}. In this situation the degrees-of-freedom are fixed and
abundance depends on the mass and energy density of sterile neutrinos.
Note that the above relation between thermal and sterile neutrino masses
is valid only for the so-called non-resonant production mechanism \citep{Dodelson:1994}. When
other mechanisms are involved \citep[e.g. resonant production][]{Boyarsky:2008a} the
relation is non-trivial.

%%%%%%%%%%%%%%%%%%%%%%%%%%%%%%%%%%%%%%%
\subsection{The Non-linear Power Spectrum}

Now that we have shown the effects of WDM on the linear matter density
field we must outline some tools for the standard model of structure
formation. In the matter dominated era, the density contrast grows and
eventually reaches unity, where non-linearities must be properly
accounted for and modelled. From this point on, standard perturbation
theory is no longer appropriate and we must employ approximation
methods as exact solutions to the Einstein field equations no longer
exist.

It is necessary to have a robust model of non-linear structure in order to take full advantage of future weak lensing data. For this reason we compare the non-linear matter power spectra extracted from simulations with  derived non-linear models. The halo model of non-linear structure is based on the assumption that large scale structure is made up of individual objects occupying peaks in the matter over-density field \citep{Press:1974,Seljak:2000,Cooray:2002}. The most important elements of this model, the mass function, the halo bias \citep{Press:1974} and the halo density profile \citep{Navarro:1997} are based on the assumptions that all dark matter in the universe is found in haloes and that there is no observable suppression of small scale over-densities from early-times free-streaming of dark matter particles or late-times thermal velocities. 

These are characteristic properties of CDM, but do not apply to WDM. For this reason one should re-visit the modelling of cosmological structure. \citet{Smith:2011,Schneider:2012,Dunstan:2011,Schneider:2013,Angulo:2013} modified the halo model by applying a specific prescription to the non-linear contribution, in addition to suppressing the initial density field, modelled by applying a transfer function from \autoreft{eq:linearpswdm} to the linear matter power spectrum. We discuss the halo model in \autoref{sec:halo}. However, we wish to first summarise some results from N-body simulations.

%%%%%%%%%%%%%%%%%%%%%%%%%%
\subsubsection{WDM Simulations}

N-body simulations have long been considered important in calculating
the properties of the large scale structure of DM and the formation of
this structure in the CDM scenario
\citep{Appel:1985,Barnes:1986,Katz:1996,Frigo:1999,Bagla:2003,Springel:2005}.
In recent years, this method has also been applied to the WDM case
\citep{Boehm:2005,Boyanovsky:2007,Wang:2007,Zavala:2009,Colombi:2009,Viel:2012,Schneider:2012,Dunstan:2011,Benson:2012,Angulo:2013,Semenov:2013}.
Numerical convergence for WDM (and HDM) simulations is particularly
difficult to achieve as pointed out by \citet{Wang:2007}. The reason
is due to the spurious fragmentation of filaments that give rise to a
pattern of small mass haloes. This effect can be alleviated by
increasing the number of particles (although convergence is slow) or
by preventing the fragmentation of such structures.  In any case,
convergence tests of the relevant simulated physical quantities (like
the \Lya forest flux and/or the mass function or matter power
spectrum) must be performed in order not to be affected by this at the
scales or redshifts of interest. A \textit{post hoc} solution was
proposed by \citet{Schneider:2013}, which does not solve the problem,
but corrects the result via subtraction of spurious haloes, while
\citet{Lovell:2013} identifies spurious haloes in the initial
conditions.
\begin{figure*}[!t]
	\centering
	\includegraphics[width=\textwidth]{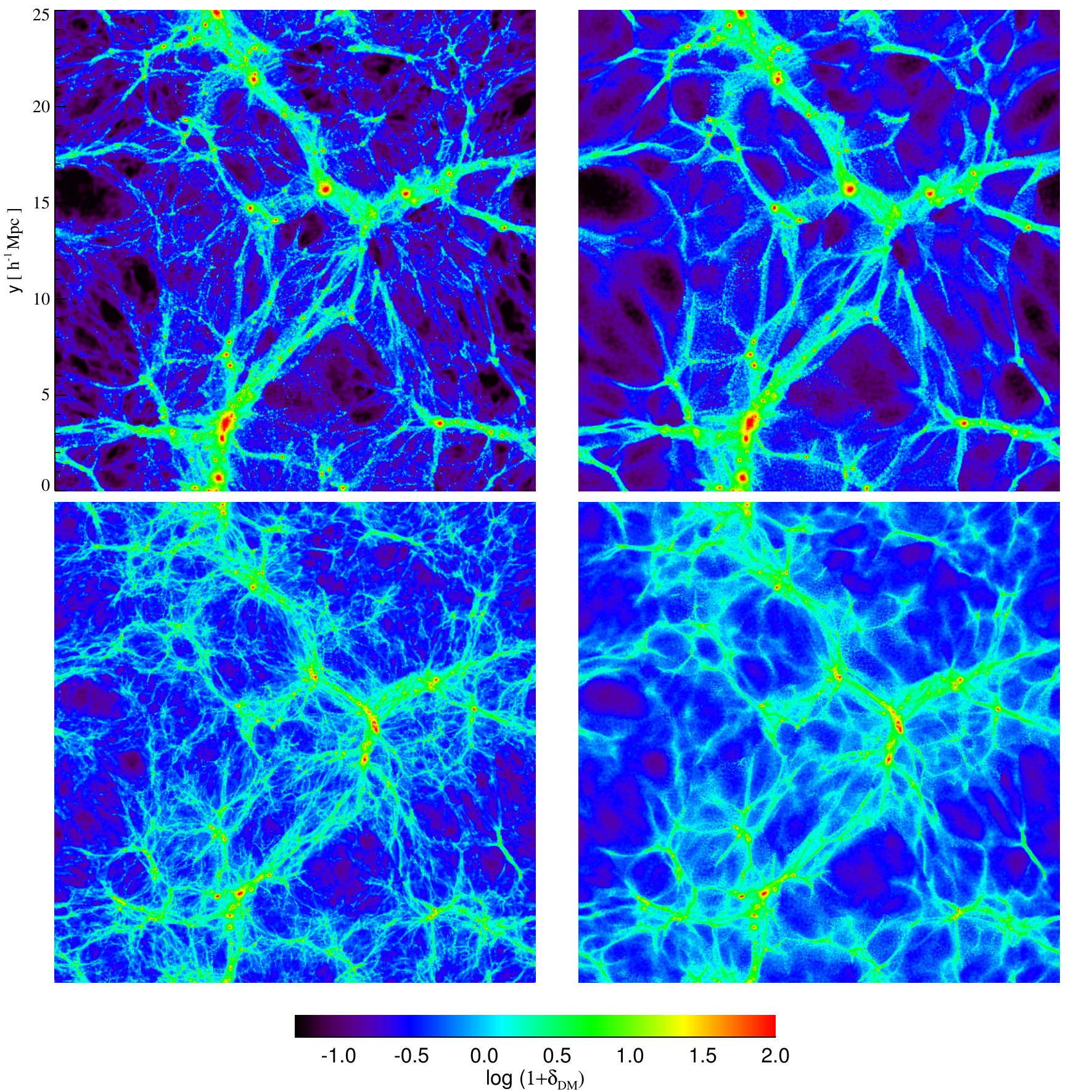}
	\caption{We plot the projected DM overdensity from a
          high-resolution hydro-dynamical simulation of
          \citet{Viel:2012}, at redshifts $z=0$ (upper panels) and
          $z=2$ (lower panels) for \LCDM and $1\keV$ WDM, in the left
          and right columns, respectively. The amount of substructure
          present in the \LCDM model is more pronounced w.r.t. the WDM
          one. The box size is 25 comoving Mpc$/h$ and the thickness
          is 5 comoving Mpc$/h$.}
\end{figure*}

N-body simulations assume that collapsing matter is non-relativistic ($\rho\gg P$) and that collapse is only possible on sub-horizon scales ($k\gg aH$). Therefore in \LCDM, the non-relativistic, Newtonian perturbation equations are sufficient and so, the collisionless Boltzmann and Poisson equation are solved in a discreet way.

These equations are normally solved by an N-body code, e.g. \code{Gadget} \citep{Springel:2000}. It is difficult to achieve this simply with finite difference methods, so 
\redtxt{Monte-Carlo-like}
N-body simulations are employed to integrate the Boltzmann equations of N particles populating the phase space, using the method of characteristics\footnote{The method of characteristics is a way of solving partial differential equations by reducing them to a set of ordinary differential equations and integrating from a set of initial conditions. In other words the partial differential equations are solved by integration along characteristic curves, in this case the characteristic curves of the collisionless Boltzmann equation.}.

\citet{Smith:2002} compared the standard CDM halo model to CDM simulations of large scale structure formation and developed an analytical fit to the non-linear corrections of the matter power spectrum, known as \code{halofit}. \citet{Markovic:2010} and \citet{Viel:2012} applied these corrections to a linear matter power suppressed by the \citet{Viel:2005} WDM transfer function \autorefp{eq:linearpswdm}. \citet{Viel:2012} ran cosmological N-body simulations (DM only) in the WDM scenario. They found that the WDM halo model is closest to simulations at $z=1$ for $1\keV$ WDM, but that it over-estimates the suppression effect at $z=0.5$ for $0.5\keV$ WDM by about $5\%$ on scales $k>1\ihMpc$. On scales $k<1\ihMpc$ however, the \code{halofit} non-linear correction describes the simulations better than the halo model, even though on smaller scales it severely underestimates the suppression effect, which becomes worse at lower redshifts. A further small modification of the WDM halo model improves its correspondence to the simulations and allows one to use it at small scales \autorefp{sec:halo}. 

\begin{figure*}[!ht]
	\centering
	\includegraphics[width=0.8\textwidth]{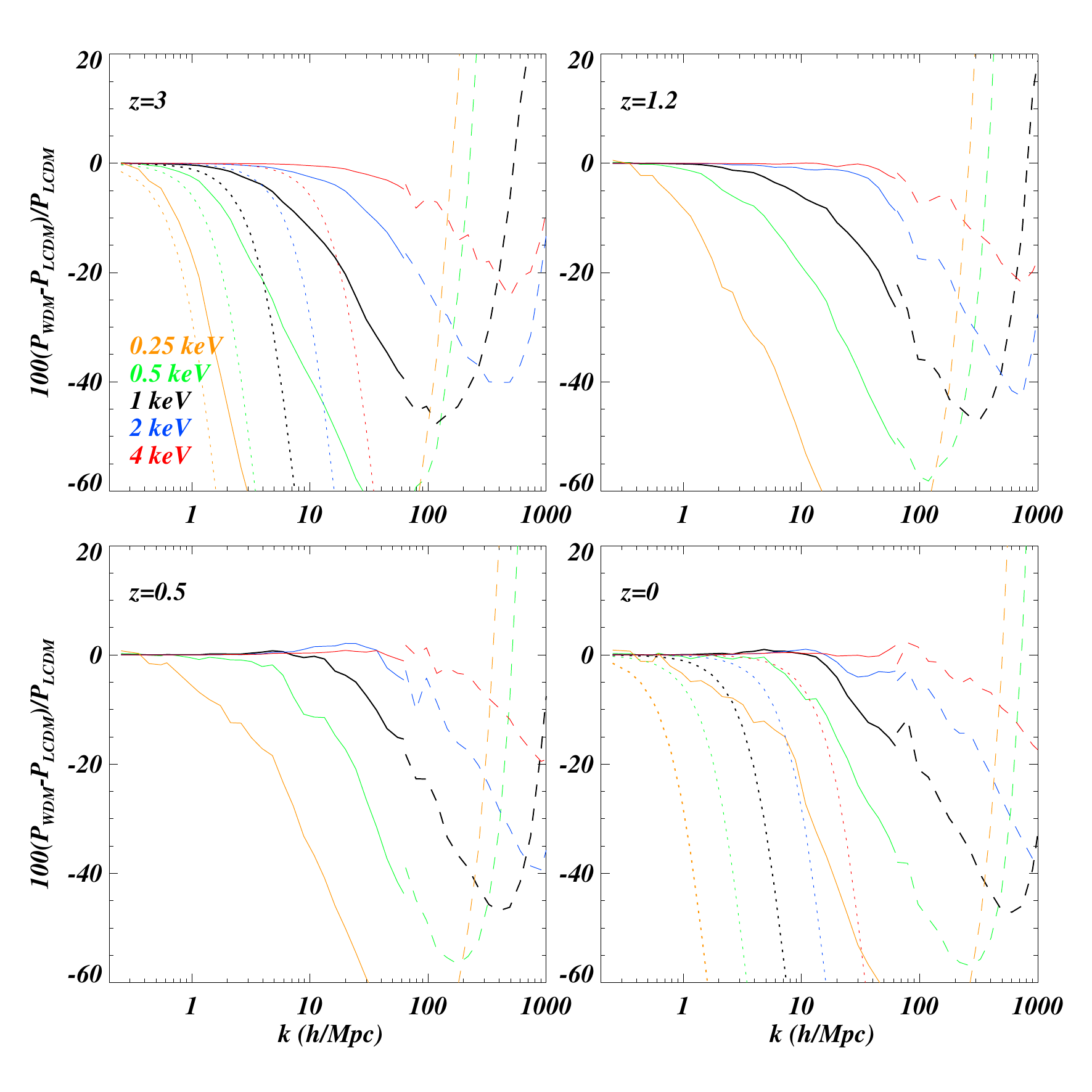}
	\caption{Percentage differences between WDM and CDM non-linear matter power spectra from hydro-dynamical simulations at high-resolution. The solid lines show the large scale power, while the dashed lines describe the small scale power obtained with the folding method in order to reach smaller scales \citep[see][for details]{Jenkins:1998,Colombi:2009}. The dotted line is the suppression to the linear matter power spectrum and is the same both in the $z=3$ and $z=0$ panels. The different panels show $z=0,0.5,1.2,3$.
	\redtxt{Note that the steep rise on scales, $k>50 \, h/Mpc$ is affected by the poor resolution of the WDM simulations and it is not fully physical (although an increase of power could be expected and it might be due to the different dark matter density profile at small scales).}
	}\label{fig:sims}
\end{figure*}

\citet{Viel:2012} consider varying resolutions and WDM models. These simulations were run using the N-body code \code{Gadget-2}, for which the initial conditions were generated using the WDM-suppressed linear matter power spectrum in \autoreft{eq:linearpswdm}. In \autoreft{fig:sims} we see plotted the percentage differences between the WDM and CDM non-linear matter power spectra for several different WDM models, denoted by the different thermal relic particle masses. This plot shows the suppression effect growing with decreasing particle mass (i.e. increasing WDM temperature), but also with increasing redshift and demonstrates the
effect that the WDM signal is erased with time due to the
non-linear growth of structure.
It may be noted that the free-streaming scale below which the power
spectrum becomes exponentially suppressed is of the order of $k \sim
1 \ihMpc$ for $1\keV$ WDM particles. The authors found a fitting
function that can be used to calculate the non-linear matter power
spectrum in the WDM scenario from the CDM $P^\nl(k)$ analogously to
\autoreft{eq:linearpswdm}: 
\be \label{redeq:fit} 
\mathcal{T}^2_\nl(k) \equiv P_\WDM(k)/P_\CDM(k)=(1+(\alpha\,k)^{\nu l})^{-s/\nu} \comma
\ee
where
\be
\alpha(m_\WDM,z) = 0.0476
	\left(\frac{1\keV}{m_\WDM}\right)^{1.85}
	\left(\frac{1+z}{2}\right)^{1.3} \comma 
\ee 
and $\nu=3$, $l=0.6$ and
$s=0.4$ are the fitting parameters. This function is applied by first
calculating the non-linear matter power spectrum using \LCDM parameter
(e.g. from {\small CAMB}) and then multiplying by the square of the
WDM ``transfer function''.

Assuming WDM to be thermal relic fermions, their relic velocities have a Fermi-Dirac distribution, which can be added to the proper velocities calculated from the gravitational potentials from linear theory. The velocities for some of the WDM models 
\redtxt{they use}
can be found to be: $v_{\rm th} \in \{27.9,11.5,4.4,1.7,0.7\} \kms$ for $m_\WDM \in \{0.25,0.5,1,2,4\}\keV$, respectively. For comparison, the typical r.m.s. value for the velocity in a \LCDM run is $v_{\rm g} \sim 30 \kms$, so it is significantly larger than any thermal velocities of WDM particles in the models that are still allowed by for example the \Lya forest ($m_\WDM \gtrsim 2\keV$).

It has been shown by many authors
\citep[e.g.][]{Schaye:2010,vanDaalen:2011,Casarini:2011,Semboloni:2011}
that baryons, making up 17\% of the total matter density, affect the
distribution of dark matter on small scales significantly. Simple
hydro-dynamical simulations in WDM and \LCDM were run by
\citet{Viel:2012}.  \redtxt{They include a prescription for radiative
  cooling and heating, where all the cooling comes from Hydrogen and
  Helium \citep[as in]{Katz:1996} and no metal cooling is considered.}
The prescription for modelling the cooling and the star formation
criterion are described in more detail in \citet{Viel:2004} and is
called ``quick \Lya'', since it can be used in order to speed up the
hydro-dynamics with practically no impact on the \Lya forest flux
statistics (this simulation is labelled as ``BARYONS+QLYA'').  A
further simulation has also been run that uses a more refined star
formation criterion and strong galactic winds powered by the thermal
feedback of supernovae (this simulation is labelled as
``BARYONS+SF+WINDS'').
\begin{figure*}[!ht]
	\centering
	\includegraphics[width=0.8\textwidth]{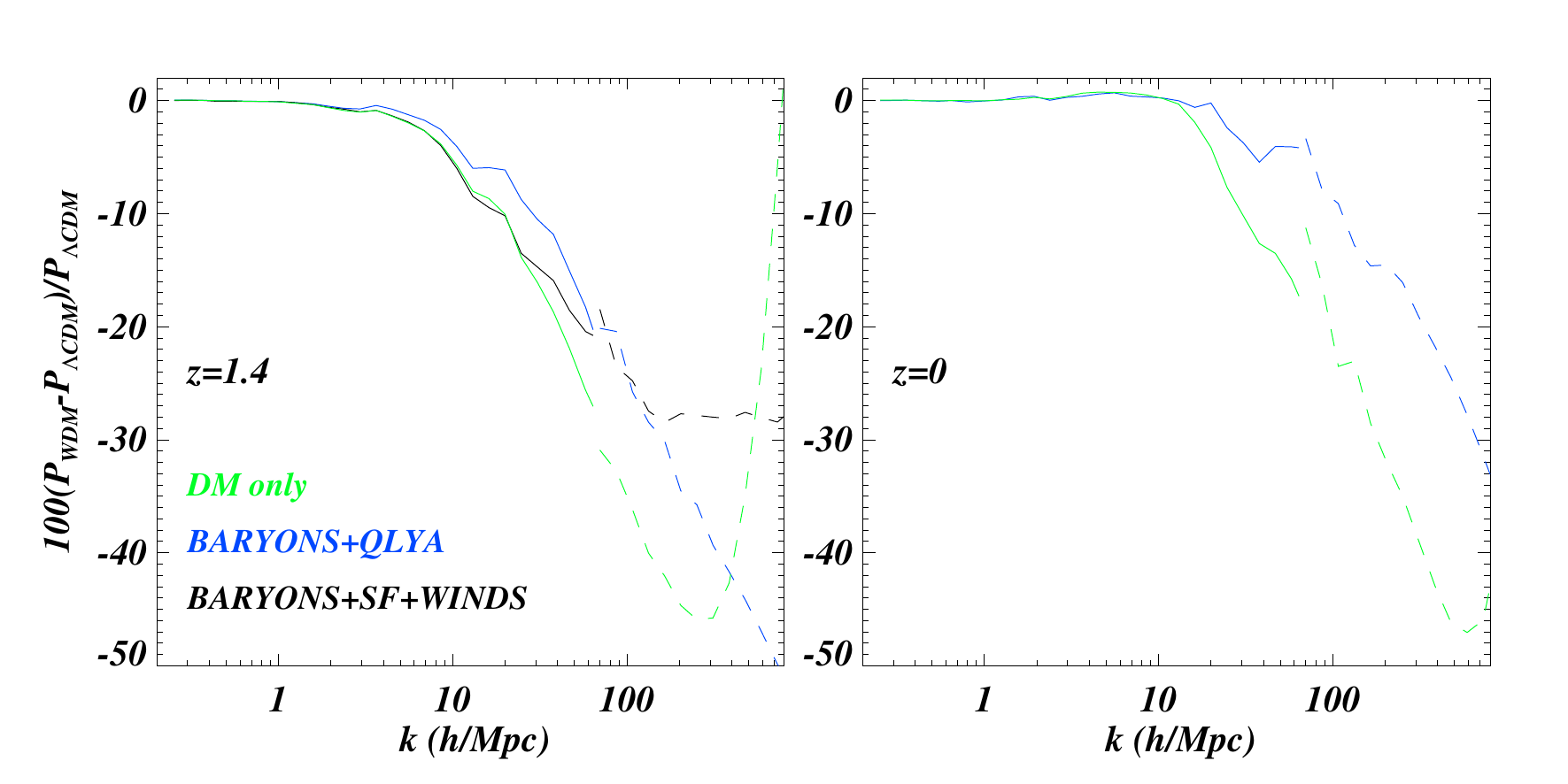}
	\caption{WDM suppression for three different simulations with
          and without baryons. These three different simulations are
          compared with the corresponding \LCDM run with the same
          initial conditions. DMONLY is the resulting percentage
          difference between the WDM and CDM non-linear matter power
          spectrum (green), BARYONS+QLYA includes cooling due to H and
          He (blue), and BARYONS+SF+WINDS, which includes star
          formation and strong galactic winds (black). The
          prescription used for the star formation processes is
          labelled as ``quick \Lya'' from
          \cite{Viel:2004}. We show two different redshifts: $z=1.4$ and $z=0$.} \label{fig:baryons}
\end{figure*}

We plot in \autoreft{fig:baryons}, the resulting percentage difference between a $1\keV$ WDM and CDM non-linear matter power spectrum, where both come from simulations that include cooling and heating processes from the ultraviolet background and a simple star formation criterion. 
\redtxt{Out}
of these simulations, one included galactic winds had to be stopped at $z = 1.2$ due to limited computational resources and is therefore plotted above this redshift. It can be seen in \autoreft{fig:baryons} that the inclusion of baryonic processes can have a very significant scale and redshift dependent effect on the suppression signal of WDM. It seems likely that some baryonic processes become more efficient in a collapsing over-density field that has been smoothed.  Because the baryonic processes affect the power on small scales, this can erase the suppression from WDM, which is relevant on similar scales \citep[see also][]{Gao:2007}.

It has also recently been reported by the authors of the \code{OWLS} simulations \citep[etc.]{vanDaalen:2011,Semboloni:2011} that the effects of baryonic processes, in particular the feedback from active galactic nuclei (AGN) can become dominant on scales that are significant to cosmology. This is certainly an important issue to consider in the future in order to realistically model the non-linear matter power.

%%%%%%%%%%%%%%%%%%%%%%%%%%
\subsubsection{The WDM Halo Model}\label{sec:halo}
 
It is interesting to note that even in the standard CDM scenario with WIMPy DM particles there exists a minimum free-streaming halo mass which is very low. \citet{Green:2005,Hofmann:2001,Schneider:2013} find such CDM minimum haloes have masses of $M\sim 10^{-6} M_\odot$.
In WDM models, this minimum mass is significantly larger. We explore this and other side effects of the primordial free-streaming of WDM on the properties and distribution of DM haloes in this section with reference to mostly the work of \citet{Markovic:2010,Smith:2011,Schneider:2012} and less so that of
\citet{Cooray:2000a,Cooray:2002,Abazajian:2004,Zavala:2009,Dunstan:2011,Lovell:2011,Pacucci:2013}. We do not discuss the work of \citet{Angulo:2013,Schneider:2013} here in detail, but it is worth noting that they also modified the halo mass function such that it works well in fitting the results of N-body simulations.\\

The halo model of large scale structure offers a tool to quantify the non-linear structure growth. It is based on the spherical collapse model, where the over-densities of the matter density field collapse as spherically symmetric objects. In the most rudimentary form, the halo model assumes that all matter can be found within DM haloes, which merge into larger and larger haloes with time (i.e. ``bottom-up''), stopping only around the present time, when further non-linear collapse is halted by the emergence of the Dark Energy component\footnote{\redtxt{In \LCDM, this would have happened $\sim5\times10^9$ years ago.}}.

The halo model assumes that halo positions are sampled from the linear theory matter distribution. As a result, there are two main contributions to the non-linear matter power spectrum. Firstly, the two-halo term, $P^\2H(k)$, which dominates on large scales, encodes the correlation between different haloes and is equal to the linear matter power spectrum on large scales, $P^{\lin}$. Secondly, the one-halo term, $P^\1H(k)$, refers to the correlations within a halo and therefore depends mostly on the Fourier transform of the density profile of the halo, $\tilde{\rho}(k,M,z)$. Both terms depend on the number of haloes as a function of halo mass, $dn/dM$, which can be found to a reasonable approximation using analytic arguments or more usually measured from numerical simulations. The total non-linear matter power spectrum from the halo model is a sum of the two terms:
\begin{eqnarray} \label{eq:3DmpsH}
P^{\1H}_{\nl}(k,z)&=&\frac{1}{(2\pi)^3}\int dM\frac{dn}{dM}\left[\frac{\tilde{\rho}(k,M,z)}{\rho_{\m,0}}\right]^2 \comma\\
P^{\2H}_{\nl}(k,z)&=&P^{\lin}(k,z)\left[\int dM\frac{dn}{dM}b(M,z)\frac{\tilde{\rho}(k,M,z)}{\rho_{\m,0}} \right]^2 \fullstop
\end{eqnarray}

\begin{figure}
	\centering
	\includegraphics[width=0.49\textwidth]{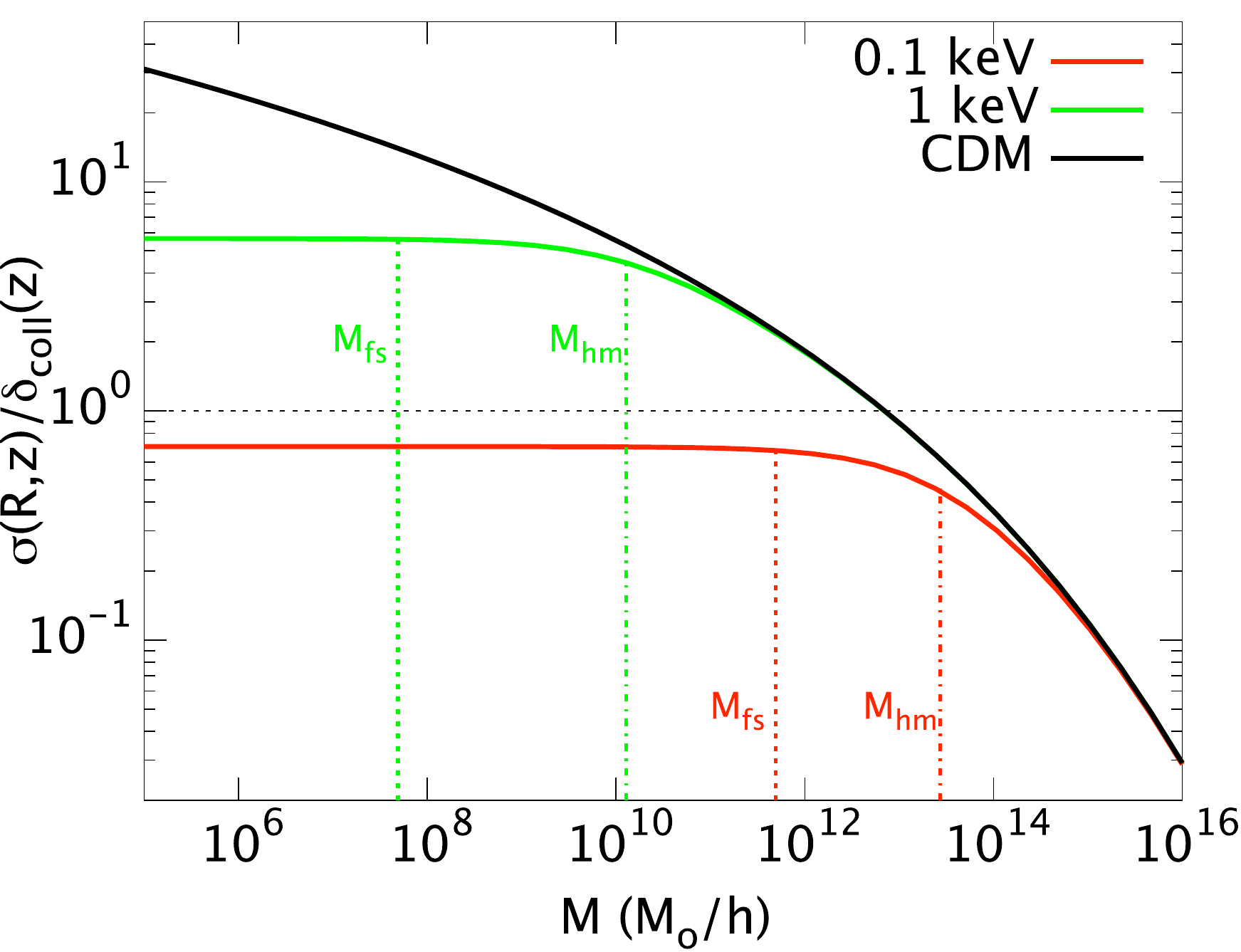}
	\caption{The root-mean-square density fluctuation for CDM (top, solid line), $0.1\keV$ WDM (bottom) and $0.25\keV$ (middle). The $\sigma(M)$ flattens off for the smallest halo masses in the WDM model, as one would expect for any smoothed field. The dotted black line indicates the critical over-density for spherical collapse.
} \label{fig:sigma}
\end{figure}
Attempts have been made by \citet{Smith:2011,Schneider:2012,Dunstan:2011} to extend the halo model to WDM scenarios by modifying its ingredients. They use the WDM linear power spectrum to calculate a new mass function using the \citet{Sheth:1999} prescription. They make the conservative assumption that the halo profiles are unchanged relative to CDM. It is in the one-halo term of the power spectrum that the effects of free-streaming of WDM are seen most strongly. This is because of the difference in the root-mean-square fluctuation, $\sigma(R)$, which becomes suppressed at small $R$ in a WDM universe. This effect is shown in \autoreft{fig:sigma} for two rather extreme WDM models, with very low particle masses. We plot this to show that for very low-mass DM particles (this is effectively HDM), the over-density field variance never reaches the necessary value for spherical collapse. This results in an extreme suppression of the formation of structure, ruling out the domination of the DM density by HDM.

\begin{figure*}
	\begin{centering}
		\includegraphics[width=0.4975\textwidth]{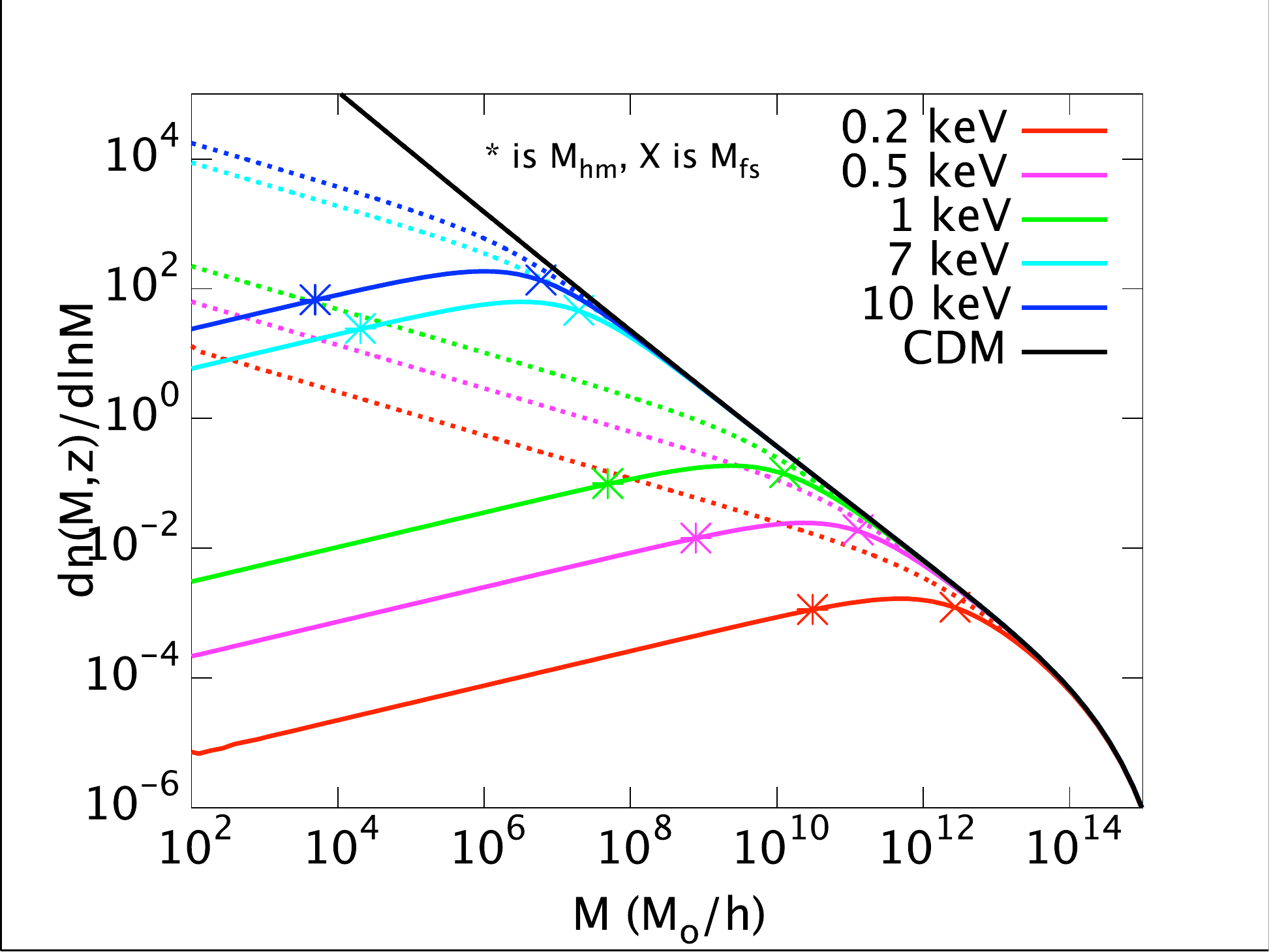}
		\includegraphics[width=0.494\textwidth]{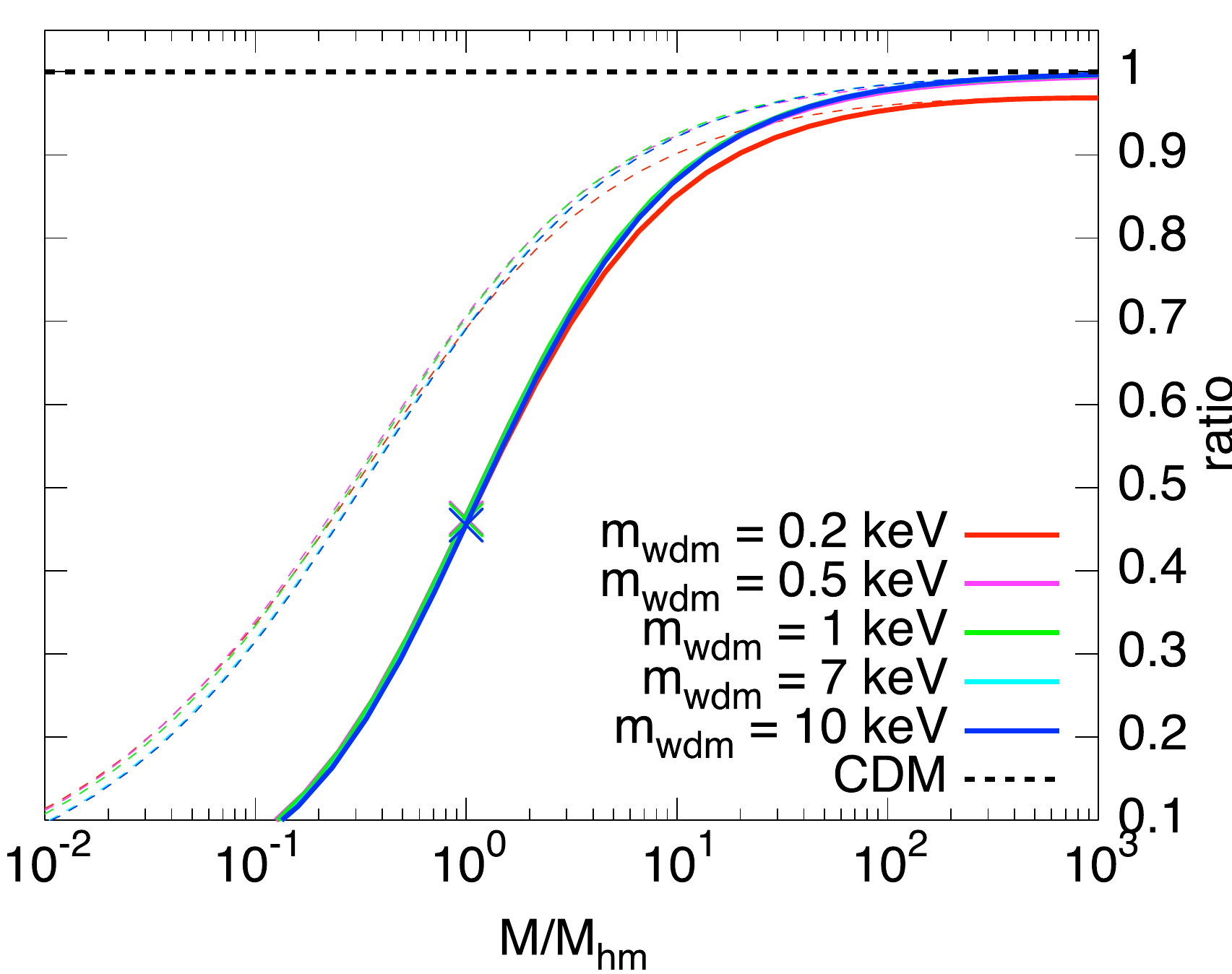}
		\caption{
		\redtxt{On the left we plot the mass functions from \cite{Sheth:1999} (dotted) vs. \citet{Schneider:2012} fit to simulations (solid).
		On the right we show the re-scaled ratios of these mass functions by the half-mode halo mass, $M_{\rm hm}$.}
The half-mode and the free-streaming halo masses are plotted with asterisks and crosses, respectively.
} \label{fig:massfns}
	\end{centering}
\end{figure*}

We explore the effect of WDM on the mass functions in \autoreft{fig:massfns}. As expected, the number density for the smallest haloes is reduced in the case of WDM. This is shown most visibly in the left panel of  \autoreft{fig:massfns}. This is useful for comparison to the general assumption of the absence of haloes below the free-streaming halo mass, defined in \autoreft{eq:fsmass} \citep{AvilaReese:2000}.
The definition of the free-streaming mass is somewhat arbitrary, because it does not really correspond to a physical halo, since it does not include the density contrast parameter, $\Delta$ \citep[as it does in][]{SommerLarsen:2001}. For this reason \citet{Schneider:2012} proposed to define instead  the half-mode halo mass, which denotes the halo mass at which the mass functions become suppressed by a factor of $1/2$:
\ba \label{eq:mhm}
\lambda_{\rm hm} &=& 2\pi \alpha \left(2^{\nu/5} - 1\right) \tab{\rm and}\nonumber\\
M_{\rm hm} &=& \frac{4\pi\bar{\rho}}{3}\left(\frac{\lambda_{\rm hm}}{2}\right)^3 \comma
\ea
where $\alpha$ comes from \autoreft{eq:scalebreak}.\\
\citet{Schneider:2012} examined the halo model in comparison to N-body simulations. They rescaled the halo masses with respect to the half-mode mass, $M_{\rm hm} \approx 2.7\times 10^3 M_\fs$, rather than the free-streaming mass as above. They find the simple fitting formula:
\be
\frac{d\tilde{n}_{\WDM}}{dn_{\WDM}} = 
	\left(1+ \frac{M}{M_{\rm hm}}\right)^{-\alpha}
	\tab , \label{blueeq:schneiderfit}
\ee
to match their simulation results well without the need to apply an artificial step function. The single fitting parameter, $\alpha = 0.6$ was able to match the simulations with less than 5\% root-mean-square error. \citet{Dunstan:2011} find very similar results.
\begin{figure}
	\centering
	\includegraphics[width=0.49\textwidth]{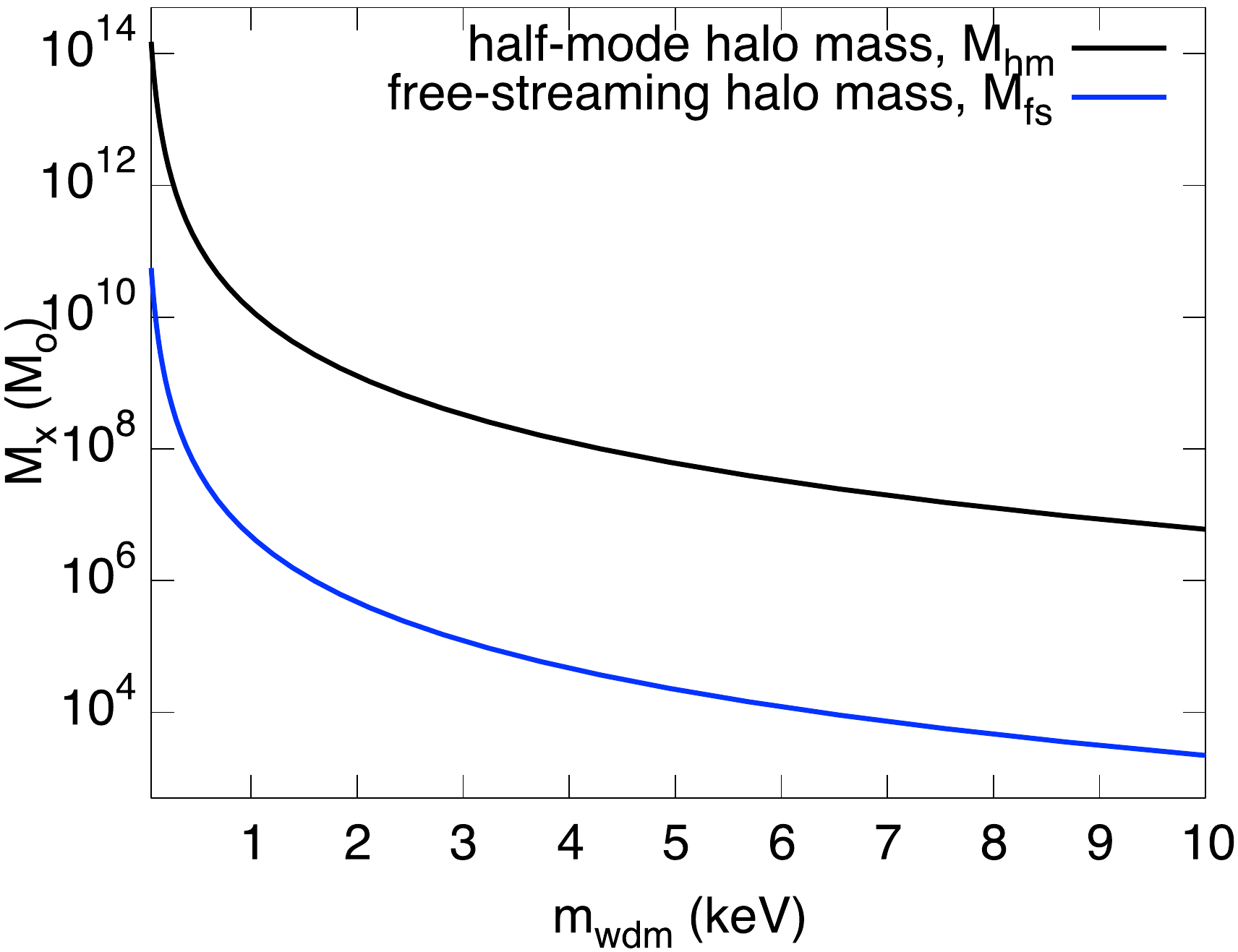}
	\caption{This figure compares the halo mass that corresponds to the ratio between WDM and CDM mass functions falling to a half, called the half-mode mass, $M_{\rm hm}$, and the free-streaming halo mass for WDM particles with masses $m_\WDM$, $M_\fs$. 
} \label{greenfig:fsmass}
\end{figure}
In \autoreft{greenfig:fsmass}, we plot the eight-times free-streaming mass against the WDM particle mass, $m_\WDM$ as well as the half-mode mass. 

This modification was motivated by the simulated mass function declining much more steeply than the \citet{Sheth:1999} already seen by \citet{Zavala:2009}. This suggests that \citet{Sheth:1999} prescription underestimates the effect of WDM on the mass functions.
On the other hand, there could be unforeseen resolution effects coming from the simulations. However this is unlikely, since there are usually spurious haloes \textit{created} in WDM simulations, which for the \citet{Zavala:2009} simulations increases the mass function $M \lesssim 10^9 \Msol$. 

\begin{figure}
	\begin{centering}
	\includegraphics[width=0.496\textwidth]{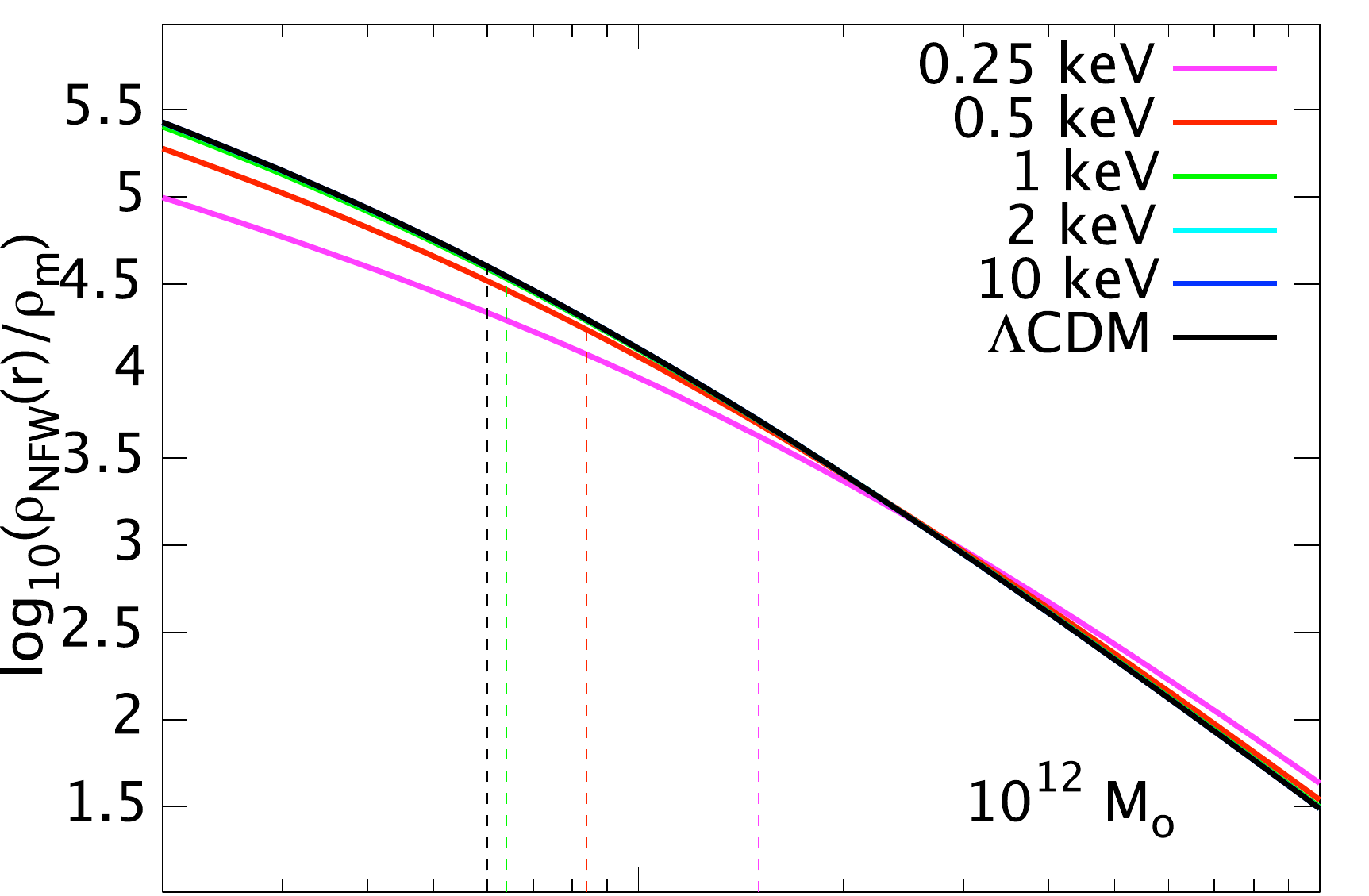}\\
	\includegraphics[width=0.496\textwidth]{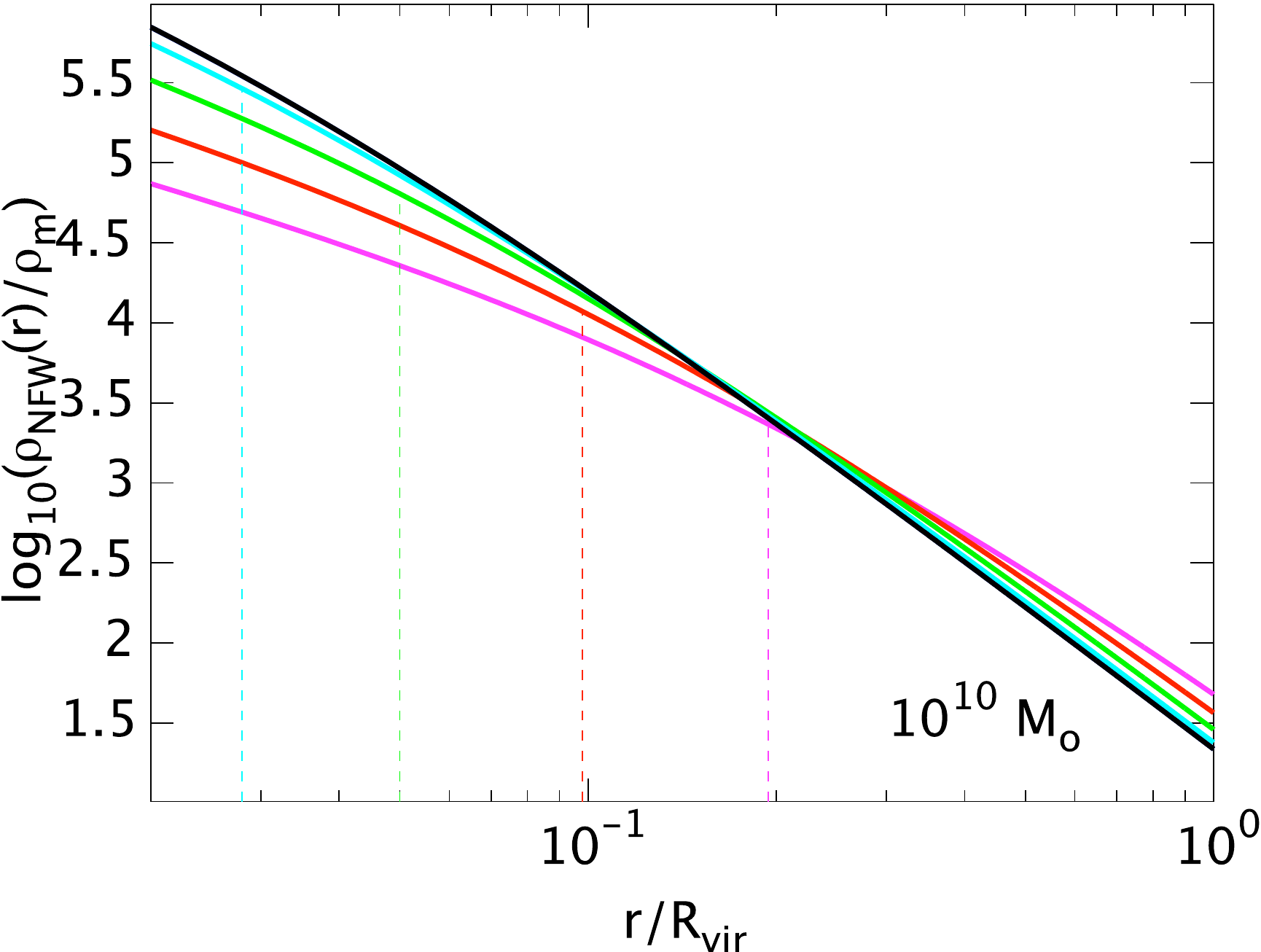}
	\caption{NFW \citep{Navarro:1997} halo density profiles for two different halo masses ($10^{10}$ and $10^{12}$ $M_\odot$) calculated with the re-scaled concentration parameter from \citet{Schneider:2012} for WDM models with $m_\WDM~\in~\{0.25,0.5,1.0,2.0,10.0\}\keV$. 
	\redtxt{The dashed vertical lines correspond to the free-streaming lengths in the different WDM models, rescaled by the virial radius of the halo.}
	} \label{fig:new}
\end{centering}
\end{figure}

Interestingly, \citet{Schneider:2012} also suggest a re-scaling of the concentration parameter to suit the simulation results better. We also plot the NFW \citep{Navarro:1997} halo density profiles calculated using the new WDM concentration parameter \citep{Seljak:2000} rescaling in \autoreft{fig:new}

\begin{figure*}[!ht]
	\begin{centering}
		\includegraphics[width=0.5\textwidth]{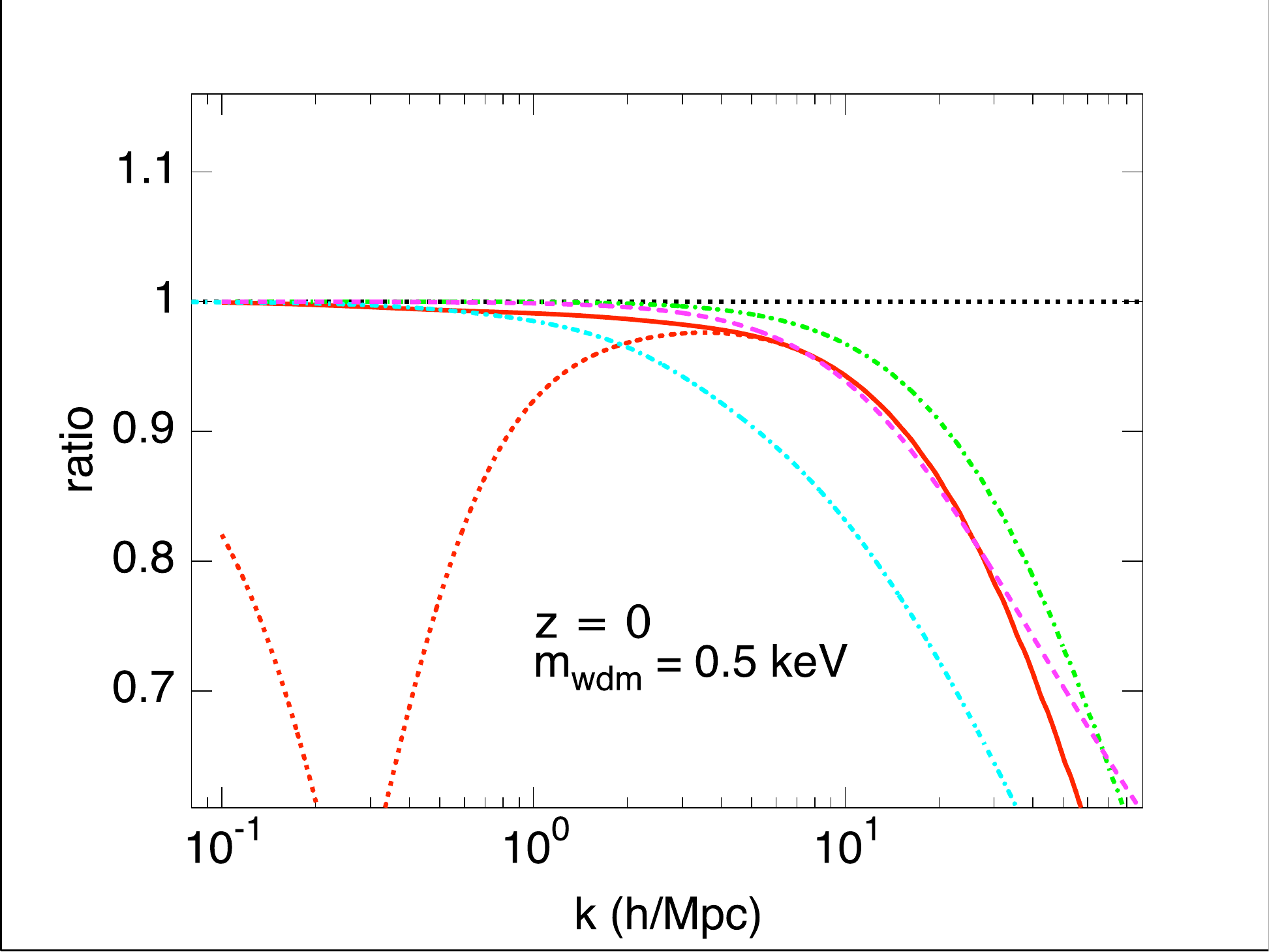}
		\includegraphics[width=0.441\textwidth]{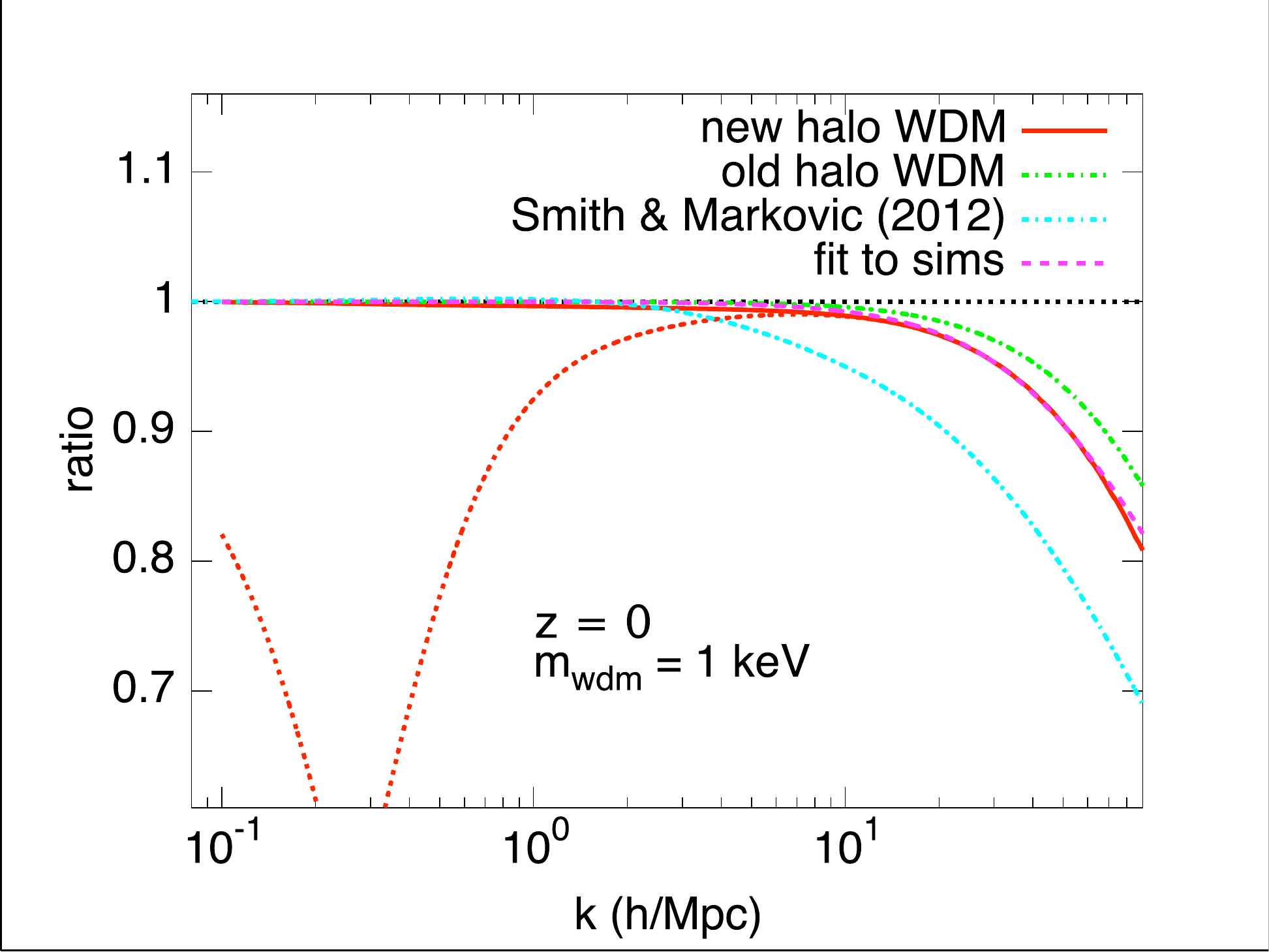}\\
		
		\includegraphics[width=0.5\textwidth]{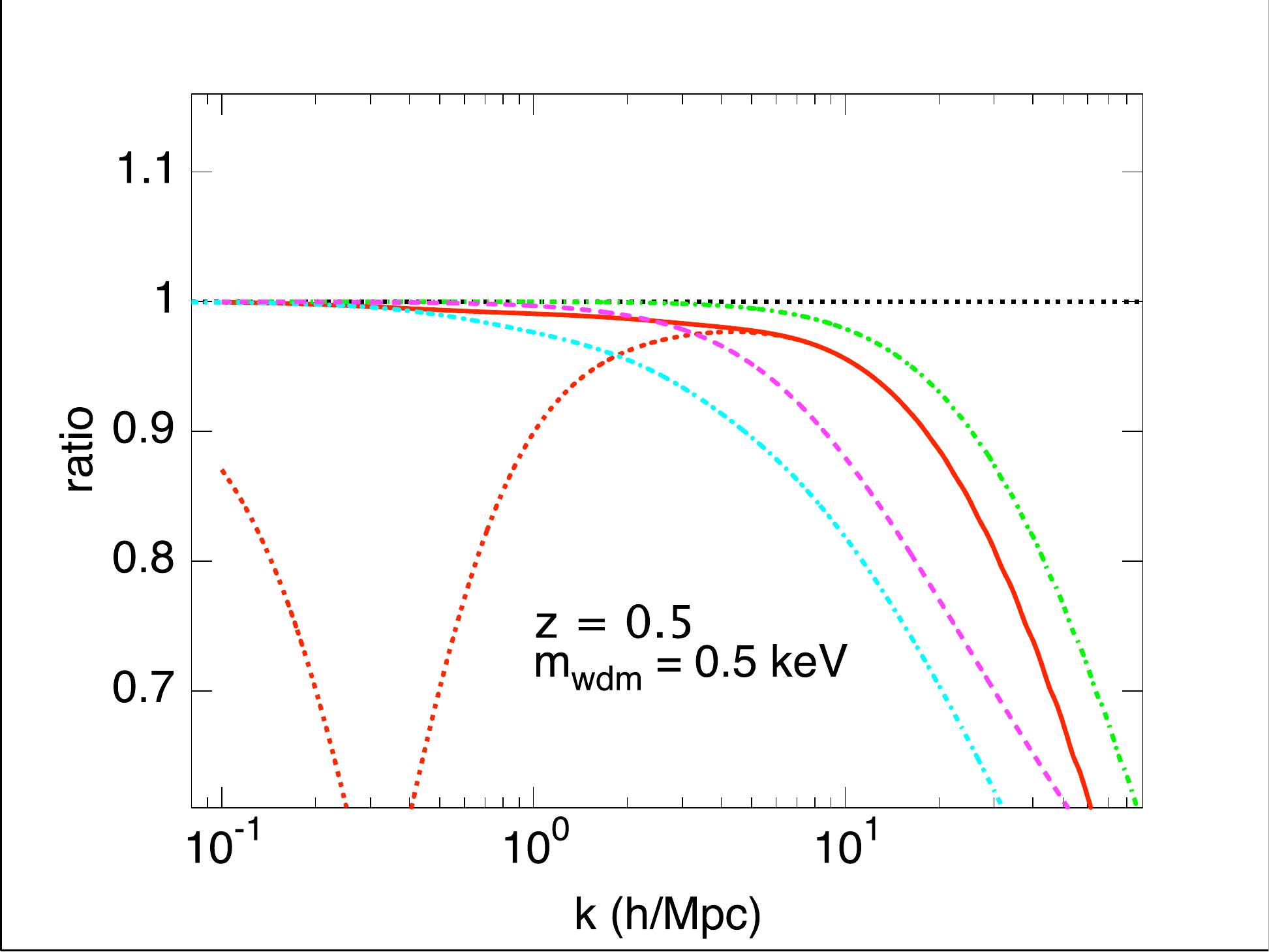}
		\includegraphics[width=0.442\textwidth]{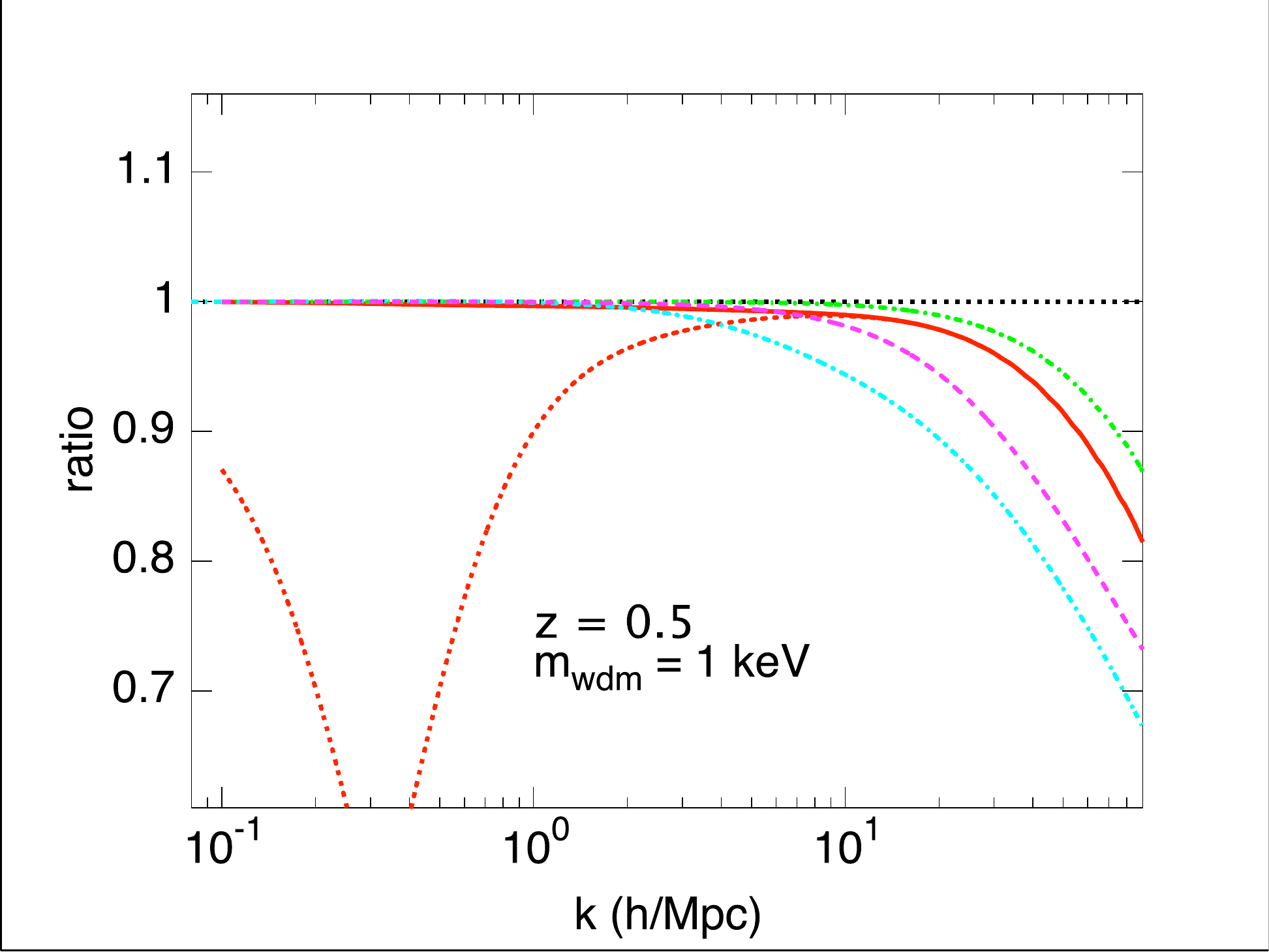}
		\caption{Ratios of non-linear matter power spectra in
                  the different models, $P^\nl_\WDM(k)/P^\nl_\CDM(k)$
                  for $500\eV$ and $1\keV$ WDM particles at $z=0$ on
                  top and $z=0.5$ at the bottom. The dotted red lines
                  show the $P^\2H(k)$ and $P^\1H(k)$ dominating at
                  small and large $k$, respectively. The red solid
                  lines show the halo model modified like in
                  \citet{Schneider:2012}. The cyan dash-dotted line
                  shows the older modification of the halo model by
                  \citet{Smith:2011}. The green dash-dotted line is
                  the simplest version of the halo model with standard
                  \citet{Sheth:1999} mass functions.
                }\label{fig:nlin}
	\end{centering}
\end{figure*}

Unfortunately, the modifications to the halo model do not seem to adequately describe the evolution of the WDM suppression with redshift. For this reason it is still the \citet{Viel:2012} fitting function for the $P^{\rm nl}_\WDM(k)$ that best fits the results from the above-mentioned simulations. We plot the ratios of the WDM vs. CDM non-linear matter power spectra in \autoreft{fig:nlin}.\\
Very recently however, \citet{Benson:2012,Schneider:2013} proposed that the mass functions should be calculated with the standard \citet{Sheth:1999} prescription, but using a sharp-k filter to find the $\sigma(R)$ instead of the real-space top-hat. This seems to describe the redshift evolution of the WDM suppression better. 
On the other hand, \citet{Pacucci:2013} propose to raise the collapse threshold to emulate the difficulty of collapse in the WDM scenario and calculate the mass functions at high redshift.

%%%%%%%%%%%%%%%%%%%%%%%%%%%%%%%%%%%%%%%
%%%%%%%%%%%%%%%%%%%%%%%%%%%%%%%%%%%%%%%
\section{Present \& Future Constraints}\label{sec:meas}

Using the methods described above, one can model the structure in the
universe and compare the resulting power spectra to observations. In
this section we will review the constraints that are obtained and
could be obtained from the two most powerful small scale observables:
the \Lya forest and weak lensing.

%%%%%%%%%%%%%%%%%%%%%%%%%%%%%%%%%%%%%%%
\subsection{The Lyman-$\alpha$ Forest}\label{sec:lya} 

%%%%%%%%%%% 
The \Lya forest, the absorption induced by intervening neutral
hydrogen along the line-of-sight to a distant quasar, is a powerful
cosmological tool ideally suited to probe the clustering of matter
over a range of scales from below the Mpc to hundreds of Mpc and from
$z=2$ up to $z=6$ \citep[for a review see][]{Meiksin:2009}. The idea behind
the cosmological applications of the \Lya forest is to relate flux
fluctuations:
\begin{equation}
\delta_{\rm F} = \frac{F - \left<F\right>}{\left<F\right>} \comma
\end{equation}
to matter fluctuations.
This can be done in several ways and  the simplest is to make
use of the so-called Fluctuating Gunn-Peterson approximation\footnote{\redtxt{This approximation neglects non-linearities induced by the fact that the signal is in redshift space.}}:

\begin{equation}
\tau_{\rm GP}=\frac{\pi\, e^2}{m_e c}f_{\alpha} \lambda_{\alpha} H^{-1}(z)\,n_{\rm HI}
\end{equation}
with $n_{\rm HI}$ 
\redtxt{being}
the neutral hydrogen density, that relates the
optical depth to the underlying density of neutral hydrogen,
$f_{\alpha}$ being the oscillator strength and
$\lambda_{\alpha}=1215.67$ \AA\, being the \Lya absorption
wavelength.  
The assumption that the gas producing the
absorption is in photoionisation equilibrium implies that $n_{\rm HI}
\propto \rho^2 T^{-0.7}/\Gamma$, where $\Gamma$ is the photoionisation
rate. Furthermore, if one assumes that the gas temperature scales as
$T=T_0(\rho/\left<\rho\right>)^\gamma$, which is set by the balance between
photo-heating and adiabatic cooling due to the expansion of the
universe and has been found to be a good approximation of the gas
thermal state at low-densities, one obtains \citep[see][]{Viel:2002}: 
\be
\tau \propto A(z) \left(\frac{\rho}{\left<\rho\right>}\right)^{\beta}
\tab {\rm with} \tab \beta=2-0.7\gamma \tab , 
\ee 
where the redshift
dependent $A$ factor will depend also on cosmological parameters,
atomic physics and on the photoionisation rate.  The observed quantity is
the transmitted flux $F=\exp({-\tau})$ and at first order it can be
easily seen that flux fluctuations are related to the linear density
contrast as $\delta_{\rm F} \propto -A\beta \delta_{\rm lin}$.
Non-linearities in the density fields and those induced by peculiar
velocities complicate the picture above and simple analytical insights
or semi-analytical models \citep{Bi:1997} must be replaced by more reliable
and accurate hydrodynamic simulations of intergalactic medium
structures performed either with smoothed-particle hydrodynamics,
Eulerian or adaptive mesh refinements codes.
%
%%------------------------------------------------------------------                                                        
\begin{table*}[!ht]
\centering 

\small
\caption{Summary of the constraints obtained on the mass of a
  WDM relic by using \Lya forest data. Apart from
  \citet{Narayanan:2000} all the other quoted values are $2\sigma$
  confidence level obtained in a Bayesian analysis.}\label{tab0}
\begin{tabular}{llcc}
Ref. & $m_\WDM$ (keV) & data & Notes \\
\hline
\hline
\noalign{\smallskip}

\citet{Narayanan:2000} & $>0.75$ &  8 high-res. &  not marginalised, N-body only\\
\citet{Viel:2005}  & $>0.55$ & 30 UVES spectra & eff.bias, hydro sims. \\
\citet{Seljak:2006b} & $>2.5$   & $\sim 3000$ SDSS spectra & approx. hydrod., full likelihood expl.\\
\citet{Viel:2006} & $>2$ &  $\sim 3000$ SDSS spectra & full hydro, approx. likelihood expl.\\
\citet{Viel:2008} & $>4.5$ & $\sim 3000$ SDSS,$\sim 60$ Keck sp. & full hydro, approx. likelihood expl.  \\
\citet{Viel:2013} & $>3.3$ & 28 high-$z$ MIKE + HIRES sp. & full hydro, good likelihood expl.\\

\hline
\noalign{\smallskip}
\end{tabular}
\end{table*}
%%--------------------------------------------------

The use of \Lya forest data to probe matter clustering has been
pioneered by Croft and co-workers at the end of the nineties: a
measurement of the linear matter power spectrum at small scales and
high redshift has been presented in \citet{Croft:2002}, by using high and
medium resolution quasar spectra together with the so-called
``effective bias'' method, $P_{\rm F}(k)=b^2_{\rm eff}(k,z)\times
P_{\rm lin}(k)$, that allowed an inversion of the one-dimensional flux
power to infer the underlying matter power spectrum. After that,
\citet{Viel:2004} used a set of about 30 high-resolution high
signal-to-noise quasar spectra taken with the UVES/VLT spectrograph
and a suite of full hydrodynamic simulation, that explored several
thermal histories, to derive the matter clustering at $z\sim 2.1$ and
$z\sim 2.7$. These data have been combined in a series of paper with
WMAP data in order to probe the long-lever arm of the matter power
spectrum and get constraints on the running of the spectral index and
inflationary parameters \citep{Viel:2006b}.  A new era in the \Lya
forest field has entered with the advent of the SDSS survey that has
allowed to obtain the 1D flux power spectrum from a set of 3000
low-resolution quasars in the range $z=2.2-4.2$ over two decades of
wavenumbers \citep{McDonald:2005} and to infer the linear matter power
spectrum amplitude, slope and curvature at $z=3$ and at a 
comoving
scale of
about $\sim$ 8 Mpc$/h$ with unprecedented precision
\citep{McDonald:2005}, by means of approximate hydro simulations.  Again
the SDSS data have been combined with other large scale structure
probes to get very tight constraints in terms of neutrino mass
fractions and cosmological parameters like running
\redtxt{of the spectral index}
 and inflation
\citep{Seljak:2006b}.  More recently, BOSS/SDSS-III has measured the
3D clustering of the flux by exploiting the signal in the transverse
direction from a set of 50000 quasar spectra: this new data set has
allowed to measure at high significance the presence of BAO (Baryonic
Acoustic Oscillations) at $z\sim 2.2$ \citep{Busca:2013,Slosar:2013} and a
new measurement of the 1D flux power has also been recently provided
\citep{Palanque:2013}.

\Lya forest data are currently providing the tightest constraints
in terms of WDM properties and there are two main reasons for this.
First of all, the one-dimensional power spectrum is a projection 
of the 3D one and contains information down to very small scales:
\begin{equation}
P_{\rm 1D,F}=\frac{1}{2\pi}\int_k^{\infty} P_{\rm 3D,F}(y)y\ dy \,
\end{equation}
and thereby is sensitive to the cutoff induced by WDM.  Secondly, \Lya
forest data span high redshift where the WDM cutoff in terms of matter
power is more pronounced and much closer to the linear behaviour (see 
\autoref{sec:frame}), in fact the \Lya forest flux power is particularly
sensitive at environments around the mean density, closer to the
linear regime and this is especially true at high redshift, due to the
strong evolution of the mean flux level. There is also another
reason that plays an important albeit minor role: the thermal
broadening depends on the temperature, which becomes colder at high
redshift, and is a fixed number in velocity space while the
free-streaming length scales as $\sqrt{1+z}$, making the thermal
contribution to a possible WDM cut-off less prominent at high
redshift.

In \autoreft{tab0} we present a summary of the constraints, in terms of the
mass of a thermal relic, that have been obtained by using \Lya forest
data.

The first constraint was obtained by \citet{Narayanan:2000}: by using N-body
simulations only and a set of eight high resolution spectra, the
 authors looked also at the flux probability distribution function and not only
at the flux power and obtain a lower limit of $0.75\keV$. The main
limitations of this work were due to the fact that no hydro
simulations were used and a proper marginalisation over nuisance
parameters was not done. \citet{Viel:2005} used instead the
effective bias method of \cite{Croft:2002} and a set of full hydro
simulations to explore the bias in WDM scenarios using 
high-resolution UVES spectra at $z=2.1,2.7$. In this case, the authors
found a 2$\sigma$ lower limit of $0.55\keV$ for a thermal relic and the
nuisance parameters were accounted for (and marginalised over) by
allowing an extra normalisation error on the data. In this paper the
authors also quote a 2$\sigma$ lower limit of $2\keV$ for a sterile
neutrino in the so-called Dodelson-Widrow scenario \citep{Dodelson:1994} and an upper limit
for a gravitino of 16 eV 2$\sigma$ C.L. in a model for which this
particle is not the total amount of DM.  Subsequently,
\citet{Seljak:2006a} exploited the unique capabilities of the
SDSS flux power spectrum of \citet{McDonald:2005} (about 3000
low-resolution low signal-to-noise QSO spectra spanning the redshift
range $z=2.2-4.2$) and showed that the constraints derived from this
data set were much tighter due to the wide redshift range probed that
allowed to break the degeneracies between cosmological and
thermal/nuisance parameters. They obtained a limit of $m_\WDM >
2.5 (14)\keV$ for a thermal relic (sterile neutrino) at the $2\sigma$
C.L. The analysis made was based on a set of approximate hydro
simulations that however explored fully the multi-dimensional
likelihood space.  The numbers derived above found confirmation in an
independent analysis of the SDSS data made by \citet{Viel:2006}
in which a suite of full hydro-dynamical simulations were used at the
expenses of a relative poor scanning of the multi-dimensional
likelihood space obtained with a Taylor expansion of the flux
power. In this work, the limits found were: $m_\WDM > 2 (10) \keV$
for a thermal relic (sterile neutrino) at the $2\sigma$ C.L., in good
agreement with the analysis of \citet{Seljak:2006a}.
\begin{figure*}[!ht]
\centering
\includegraphics[width=0.7\textwidth]{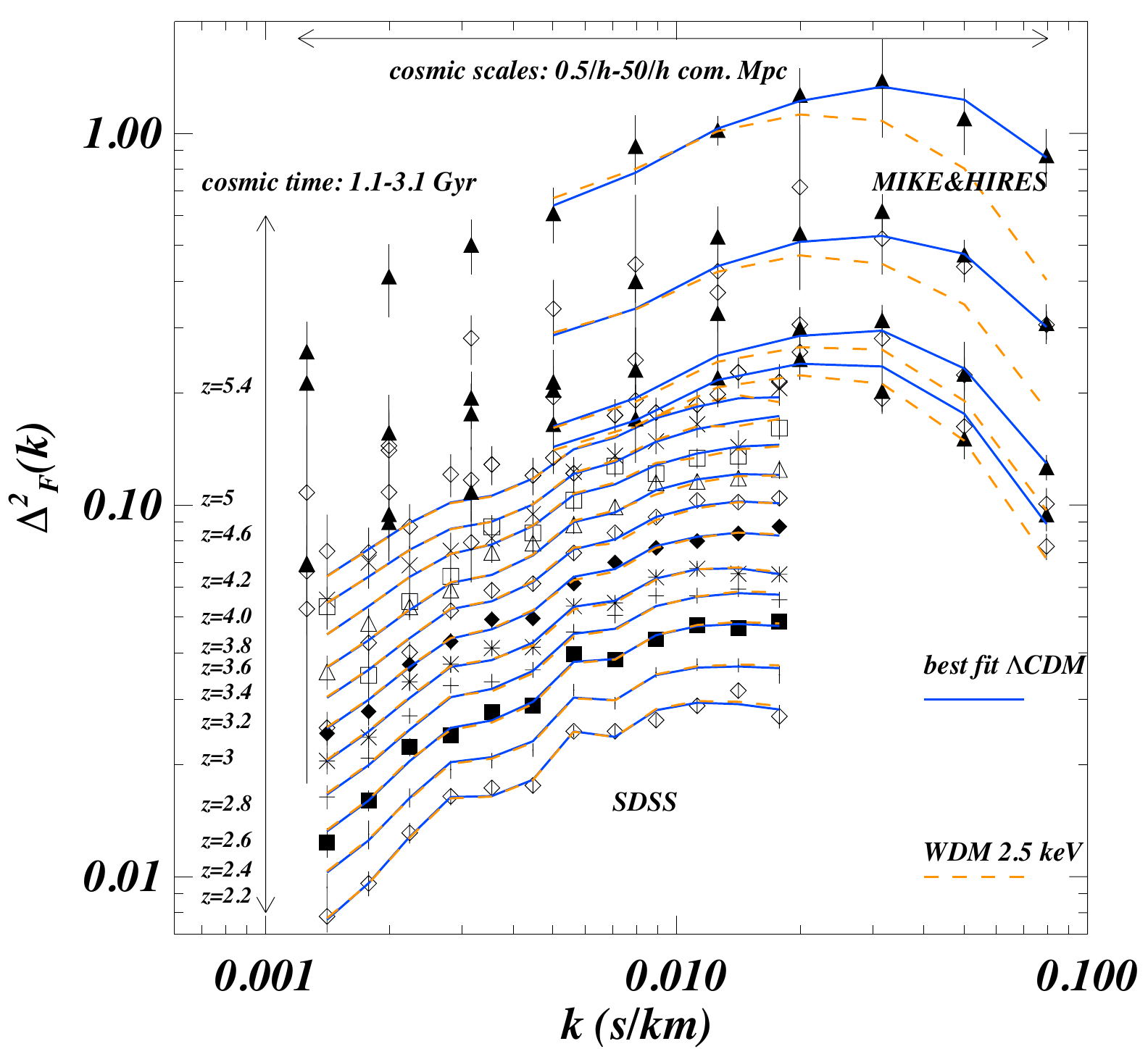}
\caption{One-dimensional flux power spectrum in dimensionless units
  ($\Delta^2(k)=P_{\rm F}(k)\times k/(2\pi)$) for the SDSS
  \citep{McDonald:2005} 
  and MIKE+HIRES \citep{Viel:2013}
   data sets. These
  data points span $z = 2.2 - 5.4$, a period of about $2\Gyr$ and about two
  decades in wavenumber space. The best fit \LCDM model is
  shown as the blue line, while the orange dashed curves are for a WDM
  model with a $m_\WDM=2.5\keV$ which is excluded by the data at
  very high significance (note that in this case the other parameters
  have been kept to their best fit values and only $m_\WDM$ is
  changed).} \label{lya}
\end{figure*}

After this \cite{Viel:2008}, explored the very high redshift regime by
using 55 high resolution Keck spectra at $z=2-6.4$ and obtained the
limits $m_\WDM > 1.2 (5.6) \keV$ for a thermal relic (sterile
neutrino) at the $2\sigma$ C.L. However, these limits greatly improved
by adding the SDSS data that allowed to break the degeneracies between
thermal and WDM cut-offs. 
\redtxt{A colder (hotter) IGM will result
in an increase (suppression) of the flux power due to the thermal
broadening of the lines, which is different: this ``thermal'' effect
could either erase or boost the WDM induced suppression.
The advantage of having a wide redshift range allows to appreciate 
the different redshift evolution of the WDM and thermal cut-offs and to lift
or break their mutual degeneracies.}
Thanks to their constraining power, these data allowed to obtain
$m_\WDM > 4.5 (28) \keV$ for a thermal relic (sterile neutrino) at the
$2\sigma$ C.L.  In this analysis a second order Taylor expansion of
the flux was used, but again the parameter space was not explored
fully and large numerical corrections were made to the flux power in
the highest redshift bins.

In \citet{Viel:2013} these numbers have been revised by using a very
comprehensive grid of hydro simulations that embrace a conservative
range of different thermal history. At these high redshift it is also
likely that galactic feedback and astrophysical effects have a much
weaker impact in terms of flux power \citep{Viel:2013b}.  In this
case the marginalisation over nuisance parameters has been made fully
in the most relevant parameter space and hydro simulations at higher
resolutions have been employed. The data used were the highest
redshift Keck spectra complemented by an equal number of MIKE
(Magellan spectrograph) at poorer resolution.  The flux power spectrum
has been measured at $z=4.2,4.6,5,5.4$ down to the scales of $k\sim
0.1$ s/km, roughly corresponding to (very non-linear) scales
$\lambda=50$ \hMpc.  The final results, that also allow for a
conservative extra error on the data side of about 30\% and is not
sensitive to continuum fitting uncertainties, give $m_\WDM > 3.3\keV$
for a thermal relic at the $2\sigma$ C.L., after having
marginalised over nuisance, ultra-violet fluctuations, instrumental
resolution, noise of the spectrograph.  From this data set and
analysis the authors concluded that thermal relics of masses $1\keV$, $2\keV$
and $2.5\keV$ are disfavoured by the data at about the $9\sigma$,
$4\sigma$ and $3\sigma$ C.L., respectively. WDM models where there is
a suppression in the linear matter power spectrum at (non-linear)
scales corresponding to $k=10\, \ihMpc$ which deviates more than 10\%
from a $\Lambda$CDM model are disfavoured by the data.  Given this
limit, the corresponding ``free-streaming mass'' below which the mass
function may be suppressed is $\sim 2\times10^8\,h^{-1}$ M$_{\odot}$.
There appears thus to be very little room for a contribution of the
free-streaming of WDM to the solution of what has been termed the
small-scale crisis of CDM.

These models have been refined further by accounting for the case of a
mixed C+WDM model in \citep{Boyarsky:2008}, where an analysis of the
SDSS and UVES data was presented. In this work the main results were
expressed in terms of a non-resonantly produced sterile neutrino and
give $m_{\rm NRP} > 8\keV$ (frequentist 99.7\% confidence limit) or
$m_{\rm NRP} > 12.1\keV$ (Bayesian 95\% credible interval) in a pure
WDM model. \redtxt{For the mixed model, they obtained limits on the mass as a
function of the WDM fraction (percentage) to be smaller than 60\% for any value
of the WDM particle mass
(frequentist 99.7\% confidence limit);
while the Bayesian joint probability allows any value of the mass (for
$m_{\rm NRP} > 5\keV$) at the 95\% confidence level, provided that the
fraction of WDM is below 35\%, for any value of the WDM particle mass. This limit can be roughly translated
into a thermal relic mass and implies that fractions of WDM below 35\%
can be accommodated only for masses above $m_\WDM>1.1\keV$.}

In \cite{Boyarsky:2008a} a mechanism of resonantly produced
sterile neutrino, that occurs in the framework of the $\nu$MSM (the
extension of the Standard Model with three right-handed neutrinos), is
analysed. Here it was shown that their cosmological signature can be
approximated by that of mixed C+WDM and for each
mass greater than or equal to $2\keV$, there exists at least one model
of sterile neutrino accounting for the totality of dark matter, and
consistent with \Lya and other cosmological data. However, the
transfer function for such candidates is quite different from the one
of the thermal relic and no direct comparison with thermal masses can
be made.

These lower limits seem to be conflicting with the upper limits
obtained on the masses of such particles coming from the observations
of the cosmic X-ray background and are: $m_{{\rm s}\nu}<1.8\keV$ at
95\% confidence \citep{Boyarsky:2006}.
\redtxt{In fact, in addition to the
dominant decay mode into three active neutrinos, the light sterile
neutrino can decay into an active one and a photon with the energy
$E_{\rm s}=m_{\rm s}/2$.  Thus, there exists a possibility of direct
detection of neutrino decay emission line from the sources with big
concentration of DM, e.g.  from the galaxy clusters
\citep{Abazajian:2001}. Similarly, the signal from radiative sterile
neutrino decays accumulated over the history of the Universe could be
seen as a feature in the diffuse extragalactic background light
spectrum. However, the constrains above assume a very simple model for
sterile neutrino production and can be circumvented by considering
other models \citep{Boyarsky:2008}.}

Overall, \Lya offers a unique probe of the matter power spectrum down
to very small scales and the tightest constraints in terms of CDM coldness.  
The most recent constraint $m_\WDM > 3.3$
keV is suggesting that the cosmic web as probed by the \Lya forest
data is quite cold and the values of WDM masses ($0.5-1.5 \keV$) that are
typically used in order to solve the missing satellite, the cusp-core
and the ``too-big-to-fail'' problems for the dynamical properties of
the most massive dwarf galaxies at low redshift are in strong tension
with the limits above.

%%%%%%%%%%%%%%%%%%%%%%%%%%%%%%%%%%%%%%%
\subsection{Cosmic Weak Lensing}\label{sec:lens}

In order to complement the \Lya constraints on the thermalised DM
particle mass, we could look at the cosmological data of sources seen
at different redshifts (tomography). An example of such a probe is
\textit{gravitational lensing}, being also the only probe that does
not rely on making assumptions about the coupling between dark and
luminous objects in that it probes directly the total gravitational
potential.

In particular, the \textit{weak gravitational lensing} induced in the
background distribution of distant galaxy images is known as
\textit{cosmic shear} and is only detectable statistically. Cosmic
shear is the weak lensing signal that is induced by the 3-dimensional
distribution of mass in the universe. We wish to describe in this
section how to theoretically calculate the weak lensing angular power
spectrum, given a 3D DM power spectrum found in the previous sections
\citep[see also][]{Bartelmann:2001}. We would like to consider
theoretical weak lensing power spectra similar to those that could be
obtained by future surveys like Euclid
\citep{Amendola:2012,Refregier:2010} and present the effect of small
scale WDM induced suppression. This will be useful for making
predictions for constraints and measurements as done in
\citet{Markovic:2010,Smith:2011,Viel:2012}.

\begin{figure*}[!ht]
\centering
\includegraphics[width=0.04\textwidth]{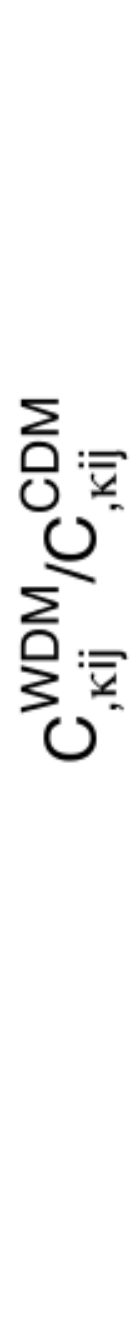}\hspace{-2pt}
\includegraphics[width=0.5\textwidth]{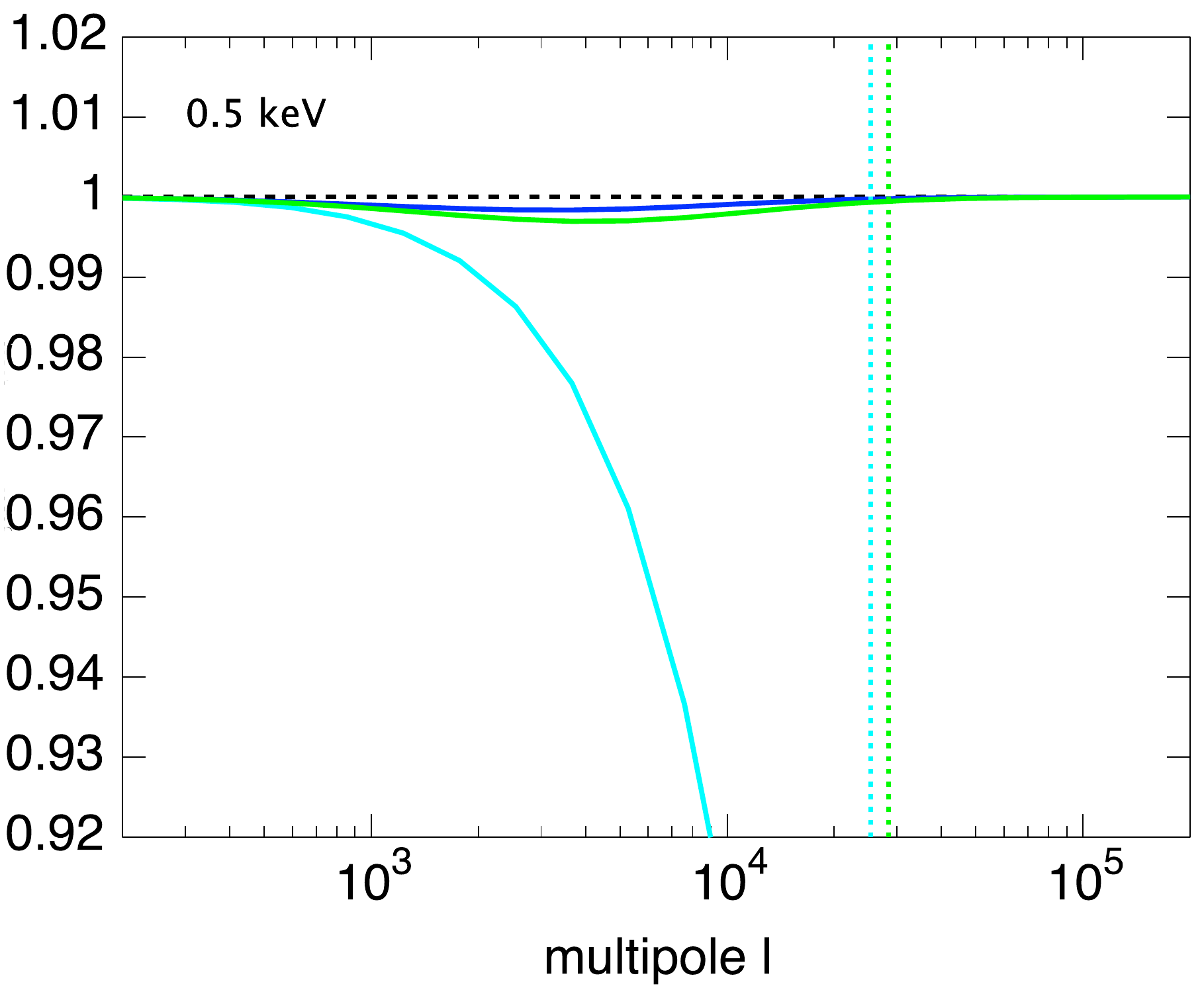}
\includegraphics[width=0.45\textwidth]{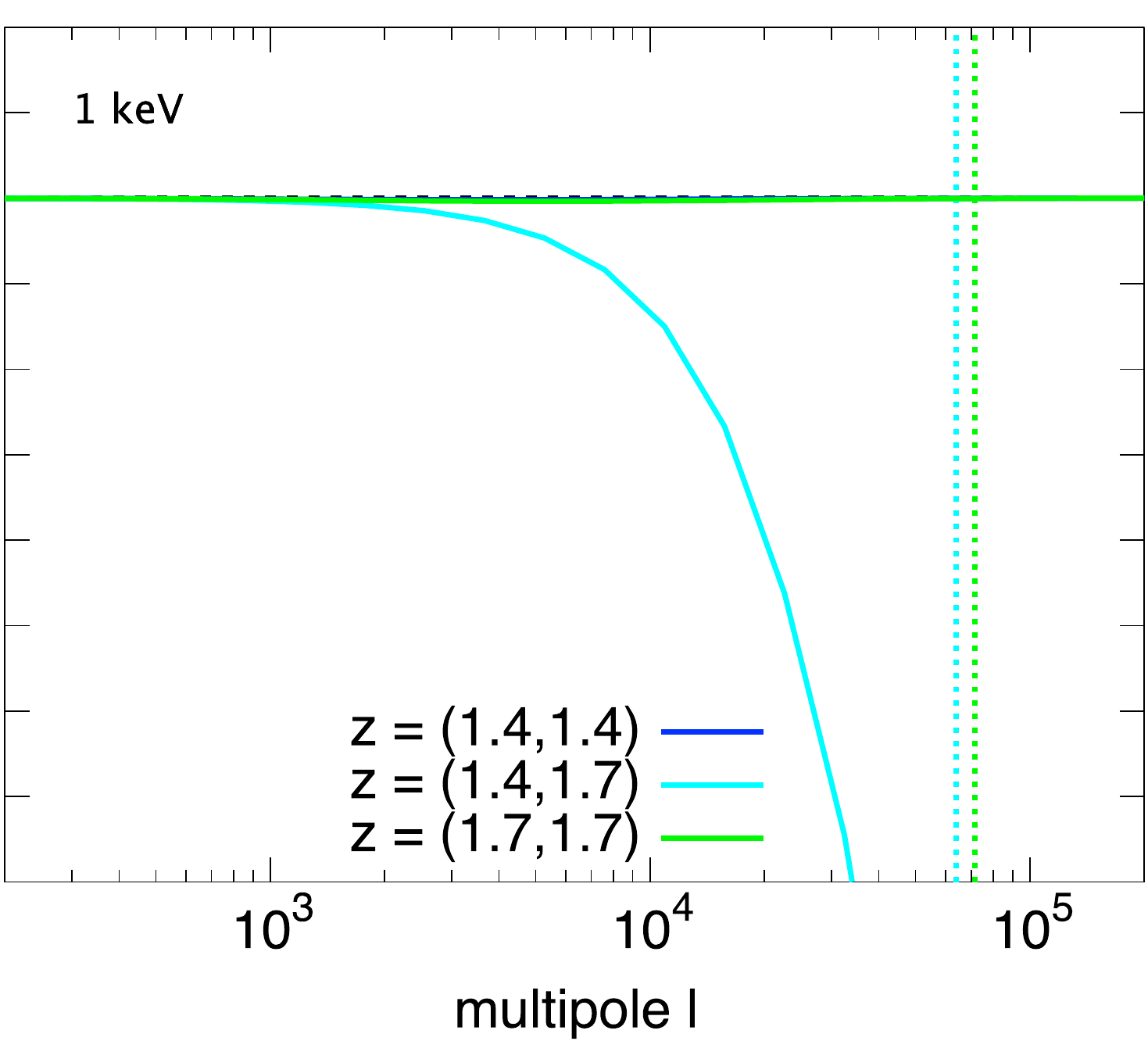}
\caption{We plot the ratios between the weak lensing power spectra obtained from a Euclid-type survey. We plot the ratios of the cross-spectra of two tomographic bins, where we have used the \citet{Viel:2012} fit for the WDM power spectrum from simulations for two different WDM models. The solid lines are the cross- and auto-correlation lensing power spectra. The dotted lines correspond approximately the $l_\fs$, i.e. the multipole value corresponding to the free-streaming scale, $k_\fs$ at the redshift or the bin (or the closer bin in the case of cross-spectra).
}\label{fig:lens}
\end{figure*}
An approximate shear power spectrum can be calculated from the halo model \citep{Cooray:2000a} and is made up of two terms, analogously to \autoreft{eq:3DmpsH}, the one-halo (or Poisson) term and the two-halo (or correlation) term
where in order to project the matter power spectrum to 2D, the small angle \citep{Limber:1953,LoVerde:2008} approximation must be made.

%%------------------------------------------------------------------                                                        
\begin{table*}[!ht]
\centering
\small
\caption{Summary of the forecasts for the Euclid survey for constraining $m_\WDM$. The lower limits are quoted to 68\% confidence, the fiducial model has been chosen as CDM (i.e. $m_\WDM \rightarrow \infty$).}\label{tab:lens}
\begin{tabular}{llcc}
Ref. & $m_\WDM$ (\keV) & model & Notes \\
\hline
\hline
\noalign{\smallskip}

\citet{Markovic:2010} & $>2.5$ & unmodified \code{Halofit} & forecast: Euclid + Planck\\
\citet{Smith:2011}  & $>2.6$ & \textit{ad hoc} WDM halo model & forecast: Euclid + Planck \\

\hline
\noalign{\smallskip}
\end{tabular}
\end{table*}
%%-------------------------------------------------- 
%
\citet{Markovic:2010,Smith:2011,Viel:2012} considered how to measure the WDM particle mass using observations of cosmic shear power spectra. From an observer's point of view, the image of each galaxy is distorted by gravitational lensing effects of all intervening matter. Therefore the cosmic shear power spectra are closely related to the matter power spectrum integrated over redshift. Future surveys are expected to use broadband photometry to estimate the redshifts of the observed galaxies. This should allow shear power spectra to be calculated different redshifts, and also allow cross power spectra between redshifts \citep[see][for a review]{Csabai:2002}.

The above-mentioned halo model approach assumes all sources at the same redshift for
simplicity, however we can expand the calculations to have a source
redshift distribution and divide the source galaxies into redshift
determined tomographic bins.  We may consider a cosmic shear survey
which has a number of galaxies per unit redshift \citep{Smail:1994}:
\begin{equation} \label{eq:nofz}
n(z)=z^\alpha e^{-(z/z_0)^\beta} \fullstop
\end{equation}

The lensing power spectra are are related to the 3D non-linear matter power spectra via:
\be\label{eq:lensing}
C_{ij}(l) = \int_{0}^{\chi_{\rm H}} d\chi_{\rm l} W_{i}(\chi_{\rm l})W_{j}(\chi_{\rm l})\chi_{\rm l}^{-2}P_{\rm nl}\left(k=\frac{l}{\chi_{\rm l}},\chi_{\rm l}\right) \ ,
\ee
where $\chi_{\rm l}(z_{\rm l})$ is the comoving distance to the lens at redshift $z_{\rm l}$ and $W_{i}$ is the lensing weight in the tomographic bin \textit{i}:
\be
W_{i}(z_{\rm l}) = \rho_{\rm m,0} \int_{z_{\rm l}}^{z_{\rm max}} \left[\frac{n_{i}(z_{\rm s})}{\Sigma_{\rm crit}(z_{\rm l},z_{\rm s})}\right] dz_{\rm s} \comma
\ee
where
\be\label{eq:sigmacrit}
\Sigma_{\rm crit}(z_{\rm l},z_{\rm s}) = \frac{c^2}{4\pi G}\frac{\chi_{\rm s}}{\chi_{\rm ls}\chi_{\rm l}}\frac{1}{(1+z_{\rm l})} \comma
\ee
and the subscripts s, l and ls denote the distance to the
source, the distance to the lens and the distance between the lens and
source, respectively.

In order to assess detectability of WDM by future weak lensing
surveys, the works above calculate predicted error bars on the weak
lensing power spectrum using the covariance matrix formalism
\citep[][]{Takada:2004} and assuming errors for a future realistic
weak lensing survey with 8 redshift bins in the range $z = 0.5 - 2.0$
(see \autoref{fig:lens}).  They additionally consider models of
non-linear WDM structure to calculate the weak lensing power
spectra. They find that for a survey like Euclid it seems to be
sufficient to model the non-linearities using the \code{Halofit}
prescription of \citet{Smith:2002}. The limits they predict for the
WDM particle mass are at the same order of magnitude as those obtained
from \Lya data \autorefp{sec:lya} and therefore, they hold the
potential to confirm the exclusion of $m_\WDM \lesssim 2\keV$. In
\autoreft{tab:lens} we quote the actual predictions made.  \redtxt{We
  note that a combination of \Lya and weak lensing can also be very
  promising in constraining the small scale clustering of matter as
  done in \citet{Lesgourgues:2007} for a standard cold dark matter scenario.}

%%%%%%%%%%%%%%%%%%%%%%%%%%%%%%%%%%%%%%%
%%%%%%%%%%%%%%%%%%%%%%%%%%%%%%%%%%%%%%%
\section{Conclusions}

This review has focussed on different approaches for modelling the
non-linear structures in our universe in the \LWDM model and on the
possibilities of constraining such a scenario \redtxt{with two
  particular sets of data: \Lya forest and cosmic shear}.  We have
made a simple choice for the WDM particle: a thermal relic with a
$\sim\keV$ mass, a mass that had been proposed in the past to solve
the so-called small scale crisis of the standard \LCDM cosmology.

\redtxt{We have decided to rely on a number of assumptions. \indent
  Firstly, in the modelling of non-linear structure, we have neglected
  the contributions of baryonic processes (although they have been
  discussed) to the shape of the potential wells of haloes. We have
  also focussed more on the clustering of dark matter and less so on
  the profiles and substructures of individual dark matter objects.
  \\ \indent Secondly, in types of observations we concentrated on
  \Lya forest and gravitational lensing and have not discussed for
  example other promising observables like the 21cm line \citep[see][]{Sitwell:2013} or the
  small-scale clustering of galaxies. \\ \indent Thirdly, we
  drastically narrowed down the range of possible particle models of
  DM, allowing only for the particle mass (and therefore temperature)
  to vary, neglecting non-neutrino-like interactions and particle
  properties. We have made these choices in order to simplify the
  analysis. Any possibilities that were neglected here were omitted
  for reasons of practicality rather than plausibility or usefulness.}

\redtxt{In summary, in order to describe a WDM} regime, it is crucial
to model the non-linearities in the matter power spectrum: thus, we
have quoted the results of N-body numerical simulations and described
the modifications to the existing halo model. In particular, we have
reported on simulations that resulted in a new WDM transfer function
for the non-linear power. This fitting function is useful for
calculating theoretical WDM power spectra and comparing them to large
scale structure data.  We have also touched upon the subject of
baryonic physics, modelled by hydro-dynamical simulations. Even though
such numerical prescriptions are yet uncertain, it is clear that
before a measurement or constraint on the WDM mass is made, one must
correctly model baryonic effects.

A possibility to disentangle cosmological
and astrophysical effects and to break this degeneracy, could be to
look at their potentially different redshift evolution. Whereas the WDM suppression increases with increasing
redshift, the effect of baryons may have an entirely different
signature. This would be important to model using
numerical methods, but it would be a large undertaking as it would
require extensive computational resources.

Another powerful tool for understanding and interpreting the large
scale structure is offered by the halo model. The halo model has now
been modified and calibrated against N-body simulations such that it
is appropriate to use in the \LWDM cosmology to predict the statistics
of the large scale structure. The new ``warm'' halo model can be
useful for a comparison with future galaxy surveys, where in order to
compare the galaxy distribution measured from observations, one must
populate the theoretical dark matter density field with galaxies. This
WDM halo model has been developed with a rather physical motivation by
\citet{Lesgourgues:2011b}
constructing the density field from dark matter haloes that host
galaxies. \redtxt{We have not discussed the clustering of galaxies
  explicitly, because competitive constraints should come only from
  small scales where the interpretation is not trivial and should rely
  on halo occupation distribution models}. Therefore we have found it
sufficient and more promising in terms of future detectability to
focus on weak lensing.

We have also shown the \citet{Schneider:2012} rescaling of the mass functions with respect to the ``half-mode mass'' and found that this resulted in power spectra that matched the N-body simulations well.
However, the lack of a prescription for how this rescaling varies with redshift, makes it more difficult to use in comparing models to data. It is likely that other prescription for the mass function \citep[e.g.][]{Angulo:2013,Pacucci:2013,Schneider:2013} will be more appropriate. Also for this reason, the fitting function to the final power spectra found by \citet{Viel:2012} describes the redshift evolution of the WDM suppression better.  

We have summarised the recent results coming from the \Lya forest data, which still provide the strongest constraints to date, of $3.3\keV$ at $2\sigma$ confidence \citep{Viel:2013}.
We have summarised forecasts for future large scale structure
surveys, in particular for the Euclid weak lensing
survey. This was done in order to show how the free-streaming
of WDM, which smoothes out the sub-$0.1\Mpc$ scales in
the linear density field impacts the present day measurements of cosmic shear. 
These forecasts have indicated that the constraints that could be placed on the $m_\WDM$ parameter, i.e. the ``warmth'' of DM, from cosmic shear will be comparable, but not stronger, than those coming from \Lya. 
In other words, because the \Lya forest probes cosmic times in the past that are much closer to the linear regime than today and is a projected measurement of the 3-dimensional density field, it has the most constraining power in measuring the small scale suppression coming from the WDM free-streaming, although the redshift range probed is very different from other observables.
  
The forecasts for Euclid show that WDM particles with masses of the order of
$m_\WDM \sim \keV$ have a large enough impact on the non-linear density
field to be detectable. It
should be noted that interesting constraints on the coldness of CDM can also be placed by using the properties of individual objects (galaxies, DM haloes, GRBs etc.) and have been presented by many authors. Future and present surveys like, Planck, \survey{SDSS} and
Euclid, \survey{SKA}, \survey{E-ELT} will also help in measuring the small scale properties of the large scale structure
and place stronger constraints in terms of the mass of the DM particle. 

\redtxt{As mentioned above, WDM is not the only model able to alter
  the \LCDM cosmology on small scales. It has become clear in the past
  decade, through the works described in this review, among others,
  that the present non-excluded models of ``standard'' WDM are on the
  limit of detectability and are, because of the increasingly tighter
  constrains on them, less able to alleviate the small-scale issues of
  \LCDM than initially hoped for. As this review was being written,
  other authors have started to come to the same conclusions
  \citep[e.g.][]{Schneider:2013b,Kennedy:2013}.  Luckily, there remain
  other types of DM models, alternative to the standard WIMP scenario,
  for example decaying or self-interacting DM that are also promising
  and worth to investigate.}

\begin{acknowledgements}
\redtxt{We thank the anonymous referee for his/her comments that helped to improve the presentation of the contents of this review.}
KM and MV would like to thank their colleagues involved in WDM topics that helped collecting the material for this review and in particular Sarah Bridle, An\v{z}e Slosar, Marco Baldi, Jochen Weller, Martin Hahnelt, George Becker, James Bolton.  KM acknowledges support from project TRR 33 `The Dark Universe'.  MV is supported by the FP7 ERC Grant ``cosmoIGM'' GA-257670 and the INFN/PD51 grant.
\end{acknowledgements}

% UNCOMMENT THE LINES BELOW IF YOU WISH TO USE BIBTEX
%\bibliographystyle{bolinx}
%\bibliography{../../../literature}

\end{document}